\documentclass[MSNbibl,seceqn,number,citesort,dvips]{arxstspdf}
\usepackage{dcolumn}
\usepackage{graphicx}
\usepackage{flushend}
\usepackage{stfloats}

\volume{28}
\issue{3}
\pubyear{2013}
\firstpage{424}
\lastpage{446}
\doi{10.1214/13-STS421} %kopijuoti is PTS
\makeatletter
\setattribute{abstract}    {skip} {25\p@}
\setattribute{keyword}     {skip} {18\p@}

\newcolumntype{d}[1]{D{.}{.}{#1}}

\newtheorem{theorem}{Theorem}[section]
\newproclaim{asp}[theorem]{Assumptions}

\newcommand{\cal}{\mathcal}
\newcommand{\uIm}{\xi^{I}_{-}}
\newcommand{\CIm}{\cC^{I}_{-}}
\newcommand{\muy}{\mu}

\newcommand{\cH}{{\cal H}}

\newcommand{\cL}{{\cal L}}

\newcommand{\cK}{{\cal K}}

\newcommand{\cC}{{\cal C}}
\newcommand{\cP}{{\cal P}}

\newcommand{\diff}{{d}}
\newcommand{\cG}{{\cal G}}
\newcommand{\Go}{\Gamma_{\mathrm{obs}}}
\newcommand{\Gd}{\Gamma_{\mathrm{db}}}
\newcommand{\qo}{q_{\mathrm{obs}}}
\newcommand{\qd}{q_{\mathrm{db}}}

\newcommand{\cN}{{\cal N}}

\newcommand{\bbE}{\mathbb{E}}
\newcommand{\bbN}{\mathbb{N}}

\newcommand{\bbI}{\mathbb{I}}

\newcommand{\bbT}{\mathbb T}

\newcommand{\bbR}{\mathbb R}

\newcommand{\R}{{\mathbb R}}

\newcommand{\bbP}{{\mathbb P}}

\newcommand{\bbZ}{\mathbb Z}

\newcommand{\etaa}{\psi}

\newcommand{\bopt}{\beta_{\mathrm{opt}}}

\makeatother

\begin{document}
\begin{frontmatter}

\title{MCMC Methods for Functions: Modifying Old Algorithms
to Make Them Faster}%\thanksref{T1}
% kai straipsnis turi susijusiu diskusiju ir rejoinder'iu
%rejoinder at \relateddoi{r}{10.1214/00-STSXXXX}.}
\runtitle{MCMC Methods for Functions}

\begin{aug}
\author[a]{\fnms{S. L.} \snm{Cotter}\corref{}\ead[label=e1]{simon.cotter@manchester.ac.uk}},
\author[b]{\fnms{G. O.} \snm{Roberts}},
\author[c]{\fnms{A. M.} \snm{Stuart}\ead[label=e2]{a.m.stuart@warwick.ac.uk}}
\and
\author[c]{\fnms{D.} \snm{White}}
\runauthor{Cotter, Roberts, Stuart and White}

\affiliation{University of Manchester, University of Warwick,
University of Warwick
and University of Warwick}

\address[a]{S. L. Cotter is Lecturer,
School of Mathematics, University of Manchester, M13 9PL, United
Kingdom \printead{e1}.}
\address[b]{G. O. Roberts is Professor, Statistics Department,
University of Warwick, Coventry, CV4 7AL, United
Kingdom.}
\address[c]{A. M. Stuart is Professor \printead{e2} and D. White is
Postdoctoral Research Assistant,
Mathematics Department, University of Warwick,
Coventry, CV4 7AL, United
Kingdom.}

\end{aug}

% ABSTRACT
%
\begin{abstract}
Many problems arising in applications result in the need to probe a
probability distribution for functions. Examples include Bayesian
nonparametric statistics and conditioned diffusion processes. Standard
MCMC algorithms typically become arbitrarily slow under the mesh
refinement dictated by nonparametric description of the unknown
function. We describe an approach to modifying a whole range of MCMC
methods, applicable whenever the target measure has density with
respect to a Gaussian process or Gaussian random field reference
measure, which ensures that their speed of convergence is robust under
mesh refinement.

Gaussian processes or random fields are fields whose marginal
distributions, when evaluated at any finite set of $N$ points, are
$\mathbb{R}^N$-valued Gaussians. The algorithmic approach that we
describe is applicable not only when the desired probability
measure has density with respect to a Gaussian process or Gaussian
random field reference measure, but also to some useful
non-Gaussian reference measures constructed through random truncation.
In the applications of interest the data is often sparse and the prior
specification is an essential part of the overall modelling strategy.
These Gaussian-based reference measures are a very flexible
modelling tool, finding wide-ranging application. Examples are shown
in density estimation, data assimilation in fluid mechanics, subsurface
geophysics and image registration.

The key design principle is to formulate the MCMC method so that it is,
in principle, applicable for functions; this may be achieved by use
of proposals based on carefully chosen time-discretizations of
stochastic dynamical systems which exactly preserve the Gaussian
reference measure. Taking this approach leads to many new algorithms
which can be implemented via minor modification of existing algorithms,
yet which show enormous speed-up on a wide range of applied problems.
\end{abstract}

% KEYWORDS
\begin{keyword}
\kwd{MCMC}
\kwd{Bayesian nonparametrics}
\kwd{algorithms}
\kwd{Gaussian random field}
\kwd{Bayesian inverse problems}
\end{keyword}
\end{frontmatter}

%s1 #&#
\section{Introduction}

The use of Gaussian process (or field) priors
is widespread in statistical
applications (geostatistics \cite{st99},
nonparametric regression \cite{hj10}, Bayesian emulator
modelling \cite{hag99}, density estimation \cite{Mac10}
and inverse quantum theory \cite{lemm03} to name but a few
substantial areas where they are commonplace).
The success of using Gaussian
priors to model an unknown function stems largely
from the model flexibility they afford, together with
recent advances in computational methodology (particularly
MCMC for exact likelihood-based methods). In this
paper we describe a wide class of statistical
problems, and an algorithmic approach to their
study, which adds to the growing literature
concerning the use of Gaussian process priors.
To be concrete, we consider a process
$\{ u(x); x\in D\}
$ for $D\subseteq\R^d$ for some $d$.
In most of the examples we consider here $u$ is not
directly observed: it is hidden (or latent) and some
complicated nonlinear function of it
generates the data at our disposal.

Gaussian processes or random fields are fields\break
whose marginal distributions, when evaluated at any finite
set of $N$ points, are $\mathbb{R}^N$-valued Gaussians.
Draws from these Gaussian probability distributions
can be computed efficiently by a variety of techniques;
for expository purposes we will focus primarily on the
use of Karhunen--Lo\'eve expansions to construct such draws, but the methods
we propose simply require the ability to draw from
Gaussian measures and the user may choose an appropriate
method for doing so. The Karhunen--Lo\'eve expansion
exploits knowledge of the eigenfunctions and eigenvalues
of the covariance operator to construct series with random
coefficients which are the desired draws;
it is introduced in Section~\ref{ssecKL}.

Gaussian processes \cite{adler2}
can be characterized by either the covariance or inverse
covariance (precision) operator.
In most statistical applications, the covariance is specified. This has
the major advantage that the
distribution can be readily marginalized to suit a prescribed
statistical use. For instance, in
geostatistics it is often enough to consider the joint distribution of
the process at locations
where data is present. However, the inverse covariance specification
has particular advantages
in the interpretability of parameters when there is information about
the local structure of
$u$. (E.g., hence the advantages of using Markov random field
models in image
analysis.) In the context where $x$ varies over a continuum (such as
ours) this creates
particular computational difficulties since we can no longer work with
a projected prior
chosen to reflect available data and quantities of interest [e.g., $\{
u(x_i); 1\le i \le m\}$ say].
Instead it is necessary to consider the entire distribution of $\{
u(x); x \in D\} $.
This poses major computational challenges, particularly in avoiding
unsatisfactory compromises
between model approximation (discretization in $x$ typically) and
computational cost.

There is a growing need in many parts of applied
mathematics to blend data with sophisticated
models involving nonlinear partial and/or stochastic
differential equations (PDEs/SDEs).
In particular, credible
mathematical models must respect physical laws and/or
Markov conditional independence relationships, which are
typically expressed through differential equations.
Gaussian priors arises naturally in this
context for several reasons. In particular:
(i) they allow for
straightforward enforcement of differentiability
properties, adapted to the model setting; and (ii)
they allow for specification of prior information in a manner
which is well-adapted to the computational tools
routinely used to solve the differential
equations themselves. Regarding (ii), it is
notable that in many applications it may be
computationally convenient to adopt an inverse covariance
(precision) operator specification, rather
than specification through the covariance function; this
allows not only specification of Markov conditional independence
relationships but also the direct use of computational tools
from numerical analysis \cite{RH05}.

This paper will consider MCMC based computational methods for
simulating from
distributions of the type described above.
Although our motivation is primarily
to nonparametric Bayesian statistical applications with
Gaussian priors, our approach
can be applied to other settings, such as conditioned diffusion
processes. Furthermore,
we also study some generalizations
of Gaussian priors which arise from truncation of
the Karhunen--Lo\'eve expansion to a random number of terms;
these can be useful to prevent overfitting and allow
the data to automatically determine the scales
about which it is informative.

Since in nonparametric Bayesian problems the unknown
of interest (a function) naturally lies in an
infinite-dimensional space, numerical schemes for\break evaluating
posterior distributions almost always rely on some kind of
finite-dimensional approximation or truncation to a
parameter space of dimension $d_{u}$, say. The Karhunen--Lo\'eve
expansion provides a natural and mathematically well-studied
approach to this problem. The larger\vadjust{\goodbreak} $d_u$ is, the better the
approximation to the infinite-dimensional \emph{true} model
becomes. However, \emph{off-the-shelf} MCMC methodology usually
suffers from a curse of dimensionality so that the numbers of
iterations required for these methods to converge diverges
with $d_u$.
Therefore, we shall aim to devise strategies which are
robust to the value of $d_u$.
Our approach will be to devise algorithms which
are well-defined mathematically for the infinite-dimensional
limit. Typically, then, finite-dimensional approximations
of such algorithms possess
robust convergence properties in terms of the choice of $d_u$.
An early specialised example of this approach within the context
of diffusions is given in~\cite{RS01}.

In practice, we shall thus demonstrate that small, but significant,
modifications of a variety of standard Markov chain Monte
Carlo (MCMC) methods
lead to substantial algorithmic speed-up when
tackling Bayesian estimation problems for functions defined
via density with respect to a Gaussian process prior,
when these problems are approximated on a finite-dimensional
space of dimension $d_u \gg1$.
Furthermore, we show that the framework adopted
encompasses a range of interesting applications.

%s1.1 #&#
\subsection{Illustration of the Key Idea}
\label{sseckey}

Crucial to our algorithm construction will be a detailed understanding
of the dominating reference Gaussian measure. Although prior
specification\break might be Gaussian, it is likely that the posterior
distribution $\mu$ is not. However, the posterior will at least be
absolutely continuous with respect to an appropriate Gaussian density.
Typically the dominating Gaussian \mbox{measure} can be chosen to be
the prior, with the corresponding Radon--Nikodym derivative just being
a re-expression of Bayes' formula
\[
{d\mu\over d\mu_0} (u) \propto{\mathsf L}(u)
\]
for likelihood ${\mathsf L}$ and Gaussian dominating measure (prior in
this case) $\mu_0$. This framework extends in a natural way to the
case where the prior
distribution is not Gaussian, but is absolutely continuous with respect
to an
appropriate Gaussian distribution.
In either case we end up with
%
%e1.1 #&#
\begin{equation}
\label{eqratz2} \frac{d\muy}{d\mu_0}(u) \propto \exp \bigl(-\Phi(u) \bigr)
\end{equation}
for some real-valued \emph{potential} $\Phi$.
We assume that $\mu_0$ is a centred Gaussian measure
$\cN(0,\cC)$.

The key algorithmic idea underlying all the algorithms
introduced in this paper is to consider
(sto\-chastic or random)\vadjust{\goodbreak} differential equations
which preserve $\mu$ or $\mu_0$ and then to employ as
proposals for Metropolis--Hastings methods
specific discretizations of these
differential equations which exactly preserve the
Gaussian reference measure $\mu_0$ when $\Phi\equiv0$;
thus, the methods do not reject in the trivial case
where $\Phi\equiv0$. This typically leads to algorithms
which are minor adjustments of well-known methods,
with major algorithmic speed-ups. We illustrate this
idea by contrasting the standard random walk method
with the pCN algorithm (studied in detail later in the paper)
which is a slight modification of the standard random walk,
and which arises from the thinking outlined above.
To this end, we define
%
%e1.2 #&#
\begin{equation}
\label{eqI} I(u)=\Phi(u)+\tfrac12\bigl\|\cC^{-1/2}u\bigr\|^2
\end{equation}
and consider the following version of
the standard random walk method:
\begin{itemize}
\item Set $k=0$ and pick $u^{(0)}$.

\item Propose $v^{(k)}=u^{(k)}+\beta\xi^{(k)},
\xi^{(k)} \sim N(0,\cC)$.

\item Set $u^{(k+1)}=v^{(k)}$ with probability
$a(u^{(k)},v^{(k)})$.
%independently of $(u^{(k)}, \xi^{(k)}).$

\item Set $u^{(k+1)}=u^{(k)}$ otherwise.

\item$k \to k+1$.
\end{itemize}

The acceptance probability is defined as
\[
a(u,v)=\min\bigl\{1,\exp \bigl(I(u)-I(v) \bigr)\bigr\}.
\]
Here, and in the next algorithm, the noise $\xi^{(k)}$
is independent of the uniform random variable used
in the accept--reject step, and this pair of random
variables is generated independently for each $k$, leading
to a Metropolis--Hastings algorithm reversible with respect
to $\mu$.

The pCN method is the following modification of the standard
random walk method:
\begin{itemize}
\item Set $k=0$ and pick $u^{(0)}$.

\item Propose $v^{(k)}=
\sqrt{(1-\beta^2)}u^{(k)}+\beta\xi^{(k)},
\xi^{(k)} \sim N(0,\cC)$.

\item Set $u^{(k+1)}=v^{(k)}$ with probability
$a(u^{(k)},v^{(k)})$.
%independently of $(u^{(k)},\xi^{(k)}).$

\item Set $u^{(k+1)}=u^{(k)}$ otherwise.

\item$k \to k+1$.
\end{itemize}

Now we set
\[
a(u,v)=\min\bigl\{1,\exp \bigl(\Phi(u)-\Phi(v) \bigr)\bigr\}.
\]

%f1 #&#
\begin{figure*}

\includegraphics{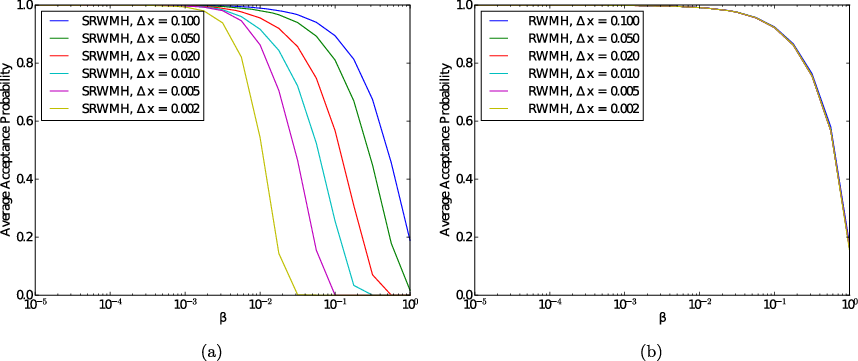}

\caption{Acceptance probabilities versus mesh-spacing, with \textup{(a)}
standard random walk and \textup{(b)} modified random walk (pCN).}\label{fig1}
\end{figure*}

The pCN method differs only slightly from the random walk
method: the proposal is not a centred random walk,
but rather of AR(1) type, and this results in a modified,
slightly simpler, acceptance probability. As is clear,
the new method accepts the proposed move with probability
one if the potential $\Phi=0$; this is because the
proposal is reversible with respect to the Gaussian
reference measure $\mu_0$.

This small change leads to significant
speed-ups for %nonparametric
problems which are discretized on a grid of dimension $d_u$. It
is then natural to compute on sequences
of problems in which the dimension $d_u$ increases,
in order to accurately sample the limiting infinite-dimensional
%nonparametric
problem. The new pCN algorithm is robust to increasing
$d_u$, whilst the standard random walk method is not.
To illustrate this idea,
we consider an example from the field of data assimilation,
introduced in detail in Section~\ref{ssecda} below,
and leading to the
need to sample measure $\mu$ of the form (\ref{eqratz2}).
In this problem $d_u=\Delta x^{-2}$, where $\Delta x$ is
the mesh-spacing used in each of the two spatial
dimensions.

Figure~\ref{fig1}(a) and (b) shows the average
acceptance probability curves, as a function of
the parameter $\beta$ appearing in the proposal,
computed by the standard
and the modified random walk (pCN) methods.
It is instructive to imagine running the algorithms
when tuned to obtain an average acceptance probability
of, say, $0.25$.
Note that for the standard method, Figure~\ref{fig1}(a), the
acceptance probability curves shift to the left as the
mesh is refined, meaning that smaller proposal variances
are required to obtain the same acceptance probability
as the mesh is refined. However, for the new method shown in
Figure~\ref{fig1}(b), the acceptance probability curves have a
limit as the mesh is refined and, hence, as the random
field model is represented more accurately; thus, a fixed
proposal variance can be used to obtain the same acceptance
probability at all levels of mesh refinement.
The practical implication of this difference in
acceptance probability curves is that
the number of steps required by the new method is
independent of the number of mesh points $d_u$ used
to represent the function, whilst for the old random
walk method it grows with $d_u$.
The new method thus mixes more rapidly than the standard
method and, furthermore, the disparity in
mixing rates becomes greater as the mesh is refined.

In this paper we demonstrate how methods such as pCN can be
derived, providing a way of thinking about algorithmic
development for Bayesian statistics which is transferable
to many different situations.
The key transferable idea is to use proposals arising
from carefully chosen discretizations of stochastic
dynamical systems which exactly preserve the
Gaussian reference measure. As demonstrated on the
example, taking this
approach leads to new algorithms which
can be implemented via minor\break modification of
existing algorithms, yet which show enormous
speed-up on a wide range of applied problems.

%s1.2 #&#
\subsection{Overview of the Paper}

Our setting is to consider measures on function
spaces which possess a density with respect to
a Gaussian random field measure, or some
related non-Gaussian measures. This setting
arises in many applications, including
the Bayesian approach to inverse problems
\cite{Stuart10} and conditioned diffusion processes
(SDEs) \cite{HSV10}.
Our goals in the paper are then fourfold:

\begin{itemize}
\item to show that a wide range of problems may be
cast in a common framework requiring samples
to be drawn from a measure known via its density
with respect to a Gaussian random field or, related, prior;

\item to explain the principles underlying
the derivation of these new MCMC algorithms for functions,
leading to desirable $d_u$-independent mixing properties;

\item to illustrate the new methods in action on some
nontrivial problems, all drawn from Bayesian nonparametric
models where inference is made concerning a function;

\item to develop some simple theoretical ideas which
give deeper understanding of the benefits of the new
methods.
\end{itemize}

Section~\ref{seccom} describes the common framework into which many
applications fit and shows a range of examples which are used
throughout the paper. Section~\ref{secprior} is concerned with the
reference (prior) measure $\mu_0$ and the assumptions that we make
about it; these assumptions form an important part of the model
specification and are guided by both modelling and implementation
issues. In Section~\ref{secmcmc} we detail the derivation of a range
of MCMC methods on function space, including generalizations of the
random walk, MALA, independence samplers, Metropolis-within-Gibbs'
samplers and the HMC\break method. We use a variety of problems to
demonstrate the new random walk method in action: Sections
\ref{ssecden2},~\ref{ssecda2},~\ref{ssecgeo2} and~\ref{ssecim2}
include examples arising from density estimation, two inverse problems
arising in oceanography and groundwater flow, and the shape
registration problem. Section~\ref{secInvP} contains a brief analysis
of these methods. We make some concluding remarks in
Section~\ref{secconc}.

Throughout
we denote by $\langle\cdot, \cdot\rangle$ the standard
Euclidean scalar product on $\mathbb{R}^m$, which
induces the
standard Euclidean norm $|\cdot|$. We also define
$\langle\cdot, \cdot
\rangle_C:= \langle C^{-{1/2}} \cdot, C^{-{1/2}}
\cdot\rangle$ for any positive-definite symmetric matrix $C$;
this induces the norm
$|\cdot|_C:=|C^{-{1/2}}\cdot|$.
Given a positive-definite self-adjoint operator $\cC$ on
a Hilbert space with inner-product
$\langle\cdot, \cdot\rangle$, we will also
define the new inner-product
$\langle\cdot, \cdot\rangle_{\cC}=
\langle\cC^{-{1/2}}\cdot, \cC^{-{1/2}} \cdot
\rangle$, with resulting norm denoted
by $\|\cdot\|_{\cC}$ or $|\cdot|_{\cC}$.

%s2 #&#
\section{Common Structure}
\label{seccom}

We will now describe a wide-ranging set of
examples which fit a common mathematical
framework giving rise to a probability measure
$\muy(du)$ on a Hilbert space $X$,\footnote{Extension to
Banach space is also possible.}
when given its density with respect to
a random field measure $\mu_0$, also on $X$.
Thus, we have the measure $\mu$ as in (\ref{eqratz2})
for some \emph{potential} $\Phi\dvtx X \to\bbR$.
We assume that $\Phi$ can be evaluated to any
desired accuracy, by means of a numerical
method. \emph{Mesh-refinement} refers to increasing
the resolution of this numerical evaluation
to obtain a desired accuracy and is tied to
the number $d_u$ of basis functions or points used
in a finite-dimensional representation of the
target function $u$.
For many problems of interest $\Phi$ satisfies
certain common properties which are detailed
in Assumptions~\ref{ass1} below. These properties
underlie much of the algorithmic development
in this paper.

A situation where (\ref{eqratz2}) arises frequently is nonparametric
density estimation (see Section~\ref{ssecden}),\break where $\mu_0$ is a
random process prior for the unnormalized log density and $\mu$ the
posterior. There are also many inverse problems in differential
equations which have this form (see Sections~\ref{ssecda},
\ref{ssecgeo} and~\ref{ssecim}). For these inverse problems we
assume that the data $y\in\bbR^{d_y}$ is obtained by applying an
operator\footnote{This operator, mapping the unknown function to
the measurement space, is sometimes termed the observation operator
in the applied literature; however, we do not use that terminology in
the paper.} $\cG$ to the unknown function $u$ and adding a realisation
of a
mean zero random variable with density $\rho$ supported on
$\bbR^{d_y}$, thereby determining $\bbP(y|u)$. That is,
%
%e2.1 #&#
\begin{equation}
\label{eqobs} y = \cG(u) + \eta,\quad \eta\sim\rho.
\end{equation}
After specifying $\mu_0(du)=\bbP(du)$,
Bayes' theorem\break gives $\muy(dy)=\bbP(u|y)$
with $\Phi(u)=-\ln\rho (y-\cG(u) )$.
We will work mainly with Gaussian random field
priors $\cN(0,\cC)$, although we will also consider
generalisations of this setting found by random truncation of
the Karhunen--Lo\'eve expansion of a Gaussian random
field. This leads to non-Gaussian priors, but
much of the methodology for the Gaussian case can be usefully
extended, as we will show.

%s2.1 #&#
\subsection{Density Estimation}
\label{ssecden}

Consider the problem of estimating the
probability density function $\rho(x)$ of a random variable
supported on $[-\ell,\ell]$, given $d_{y}$
i.i.d. observations $y_i$. To ensure positivity and
normalisation, we may write
%
%e2.2 #&#
\begin{equation}\label{eqrho}
\rho(x)=\frac{\exp (u(x) )}{\int_{-\ell}^{\ell}
\exp (u(s) )\,ds}.
\end{equation}
If we place a Gaussian process prior $\mu_0$ on $u$
and apply Bayes' theorem, then we obtain formula
(\ref{eqratz2}) with $\Phi(u)=-\sum_{i=1}^{d_y} \ln\rho(y_i)$
and $\rho$ given by (\ref{eqrho}).

%s2.2 #&#
\subsection{Data Assimilation in Fluid Mechanics}
\label{ssecda}

In weather forecasting and oceanography it is
frequently of interest to determine the initial
condition $u$ for a PDE dynamical system modelling
a fluid, given observations \cite{ben02,kal03}.
To gain insight into such problems, we consider
a model of incompressible fluid flow, namely,
either the Stokes ($\gamma=0$)
or Navier--Stokes equation ($\gamma=1$), on a two-dimen\-sional
unit torus $\bbT^2$. In the following
$v(\cdot,t)$ denotes the velocity field at time~$t$, $u$ the initial
velocity field and $p(\cdot,t)$ the pressure field at time $t$
and the following is an implicit nonlinear equation for the pair $(v,p)$:
%
%e2.3 #&#
\begin{eqnarray}
\label{NS}\qquad \partial_t v - \nu\triangle v + \gamma v \cdot\nabla v +
\nabla p &=& \etaa\nonumber\\
&&\eqntext{\forall(x,t) \in\bbT^2\times(0,\infty),}
\\[-8pt]\\[-8pt]\nonumber
\nabla\cdot v &=& 0\quad \forall t \in(0,\infty),
\\
v(x,0) &=& u(x),\quad  x \in\mathbb{T}^2.
\nonumber
\end{eqnarray}
The aim in many applications
is to determine the initial state of the fluid velocity, the function
$u$, from some observations relating to the velocity field $v$
at later times.

A simple model of the situation arising in weather
forecasting is to determine $v$ from \emph{Eulerian
data} of the form
$y = \{y_{j,k}\}_{j,k=1}^{N,M}$, where
%
%e2.4 #&#
\begin{equation}\label{eedata}
y_{j,k} \sim\mathcal{N} \bigl(v(x_j,t_k),
\Gamma \bigr).
\end{equation}
Thus, the inverse problem is to find $u$ from $y$ of the
form (\ref{eqobs}) with $\cG_{j,k}(u)=v(x_j,t_k)$.

In oceanography \emph{Lagrangian data} is often
encountered:
data is gathered from the trajectories of particles $z_j(t)$
moving in the velocity field of interest, and thus satisfying
the integral equation
%
%e2.5 #&#
\begin{equation}
\label{eqlag} z_j(t)= z_{j,0}+\int_0^t
v \bigl(z_j(s),s \bigr)\,ds.
\end{equation}
Data is of the form
%
%e2.6 #&#
\begin{equation}\label{eldata}
y_{j,k} \sim\mathcal{N} \bigl(z_j(t_k),\Gamma
\bigr).
\end{equation}
Thus, the inverse problem is to find $u$ from $y$ of the
form (\ref{eqobs}) with $\cG_{j,k}(u)=z_j(t_k)$.

%s2.3 #&#
\subsection{Groundwater Flow}
\label{ssecgeo}

In the study of groundwater flow an important
inverse problem is to determine the permeability
$k$ of the subsurface rock from measurements
of the head (water table height) $p$ \cite{mct}.
To ensure the (physically required) positivity
of $k$, we write $k(x)=\exp (u(x) )$ and
recast the inverse problem\vadjust{\goodbreak} as one for the function $u$.
The head $p$ solves the PDE
%
%e2.7 #&#
\begin{eqnarray}
\label{eqhead} -\nabla\cdot \bigl(\exp (u )\nabla p \bigr)&=&g,\quad x\in D,
\nonumber\\[-8pt]\\[-8pt]
p&=&h,\quad x \in\partial D.
\nonumber
\end{eqnarray}
Here $D$ is a domain containing the measurement points $x_i$ and
$\partial D$ its boundary; in the simplest case $g$ and $h$ are
known. The forward solution operator is $\cG(u)_j=p(x_j)$. The
inverse problem is to find $u$, given $y$ of the form
(\ref{eqobs}).

%s2.4 #&#
\subsection{Image Registration}
\label{ssecim}

In many applications arising in medicine and
security it is of interest to calculate the
distance between a curve $\Go$, given only
through a finite set of noisy observations,
and a curve $\Gd$ from a database of known
outcomes. As we demonstrate below,
this may be recast as an inverse
problem for two functions, the first, $\eta$,
representing reparameterisation of the database curve $\Gd$
and the second, $p$, representing a momentum variable,
normal to the curve $\Gd$, which initiates a
dynamical evolution of the reparameterized
curve in an attempt to match observations
of the curve $\Go$. This approach to inversion
is described in \cite{Co2008} and developed in the
Bayesian context in \cite{CC10}. Here we outline the
methodology.

Suppose for a moment that we know the entire observed
curve $\Go$ and that it is noise free.
We parameterize $\Gd$ by $\qd$ and $\Go$ by $\qo$,
$s\in[0,1]$. We wish to find a path $q(s,t)$,
$t\in[0,1]$, between $\Gd$ and $\Go$, satisfying
%
%e2.8 #&#
\begin{equation}
\label{ebcs}\qquad
q(s,0) = \qd\bigl(\eta(s)\bigr),\quad q(s,1) = \qo(s),
\end{equation}
where $\eta$ is an orientation-preserving reparameterisation.
Following the methodology of \cite{MiYo2001,GlTrYo04,VaGl2005},
we constrain the motion of the curve $q(s,t)$ by asking
that the evolution between the two curves results from the
differential equation
%
%e2.9 #&#
\begin{equation}
\label{emotion} \frac{\partial}{\partial t}q(s,t) = v \bigl(q(s,t),t \bigr).
\end{equation}
Here $v(x,t)$ is a time-parameterized family of vector fields on
$\mathbb{R}^2$ chosen as follows.
We define a metric on the ``length'' of paths as
%
%e2.10 #&#
\begin{equation}
\label{edistance} \int_0^1 \frac{1}{2}
\|v\|^2_B \,\diff{t},
\end{equation}
where $B$ is some appropriately chosen Hilbert space.
The dynamics (\ref{emotion})
are defined by choosing an appropriate $v$
which minimizes this metric, subject
to the end point constraints (\ref{ebcs}).

In \cite{Co2008} it is shown that this minimisation
problem can be solved via a dynamical system obtained
from the\vadjust{\goodbreak} Euler--Lagrange equation. This dynamical
system yields $q(s,1)=G(p,\eta,s)$, where $p$ is an initial
momentum variable normal to $\Gd$, and $\eta$ is the
reparameterisation. In the perfectly observed scenario
the optimal values of $u=(p,\eta)$ solve the equation
$G(u,s):=G(p,\eta,s)=\qo(s)$.

In the partially and noisily observed scenario we are
given observations
\begin{eqnarray*}
y_j &=& \qo(s_j)+\eta_j
\\
&=&G(u,s_j)+\eta_j
\end{eqnarray*}
for $j=1,\ldots, J$; the $\eta_j$ represent noise.
Thus, we have data in the form (\ref{eqobs})
with $\cG_j(u)=G(u,s_j)$.
The inverse problem is to find the
distributions on $p$ and $\eta$, given a prior distribution
on them, a distribution on $\eta$ and the data $y$.

%s2.5 #&#
\subsection{Conditioned Diffusions}
\label{sseccond}

The preceding examples all concern Bayesian nonparametric
formulation of inverse problems in which a Gaussian prior
is adopted. However, the methodology that we employ
readily extends to
any situation in which the target
distribution is absolutely continuous with respect to a
reference Gaussian field law, as arises
for certain conditioned diffusion processes~\cite{HSV10}.
The objective in these problems is to
find $u(t)$ solving the equation
\[
du(t)=f \bigl(u(t) \bigr)\,dt+\gamma \,dB(t),
\]
where $B$ is a Brownian motion
and where $u$ is conditioned on, for example, (i) end-point constraints
(bridge diffusions, arising in econometrics
and chemical reactions); (ii) observation of
a single sample path $y(t)$ given by
\[
dy(t)=g \bigl(u(t) \bigr)\,dt+\sigma \,dW(t)
\]
for some Brownian motion $W$ (continuous
time signal processing); or (iii) discrete
observations of the path given by
\[
y_j=h \bigl(u(t_j) \bigr)+\eta_j.
\]
For
all three problems
use of the Girsanov formula, which allows expression
of the density of the path\-space measure arising with nonzero drift
in terms of that arising with zero-drift,
enables all three problems to be written
in the form (\ref{eqratz2}).

%s3 #&#
\section{Specification of the Reference Measure}
\label{secprior}

The class of algorithms that we describe
are primarily based on measures defined through density with\vadjust{\goodbreak}
respect to random field model $\mu_0=\cN(0,\cC)$,
denoting a centred Gaussian with covariance operator $\cC$.
To be able to implement the algorithms in this
paper in an efficient way, it is necessary to make
assumptions about this Gaussian reference measure.
We assume that information about $\mu_0$ can be
obtained in at least one of the following three ways:

\begin{enumerate}
\item the eigenpairs $(\phi_i,\lambda_i^2)$
of $\cC$ are known so that exact draws from
$\mu_0$ can be made from truncation of
the Karhunen--Lo\'eve expansion and that, furthermore,
efficient methods exist for evaluation of
the resulting sum (such as the FFT);

\item exact draws from $\mu_0$ can be made
on a mesh, for example, by
building on exact sampling methods for Brownian
motion or the stationary Ornstein--Uhlenbeck (OU) process
or other simple Gaussian process priors;

\item the precision operator $\cL=\cC^{-1}$ is known
and efficient numerical methods exist for the inversion
of $(I+\zeta\cL)$ for $\zeta>0$.
\end{enumerate}

These assumptions are not mutually exclusive and
for many problems two or more of these will be possible.
Both precision and Karhunen--Lo\'eve representations link
naturally to efficient computational tools
that have been developed in numerical analysis.
Specifically, the precision operator $\cL$ is often defined
via differential operators and the operator $(I+\zeta\cL)$
can be approximated, and efficiently inverted,
by finite element or finite difference methods;
similarly, the Karhunen--Lo\'eve expansion links naturally to the
use of spectral methods.
The book \cite{RH05} describes the literature concerning
methods for sampling from Gaussian random fields,
and links with efficient numerical methods for
inversion of differential operators.
An early theoretical exploration of the links between numerical
analysis and statistics is undertaken in \cite{Dia88}.
The particular links that we develop in this
paper are not yet fully exploited
in applications and we highlight
the possibility of doing so.

%s3.1 #&#
\subsection{The Karhunen--Lo\'eve Expansion}
\label{ssecKL}

The book \cite{adler2}
introduces the Karhunen--Lo\'eve expansion
and its properties.
Let $\mu_0=\cN(0,\cC)$ denote a Gaussian measure
on a Hilbert space $X$.
Recall that
the orthonormalized eigenvalue/eigenfunction pairs of
$\cC$ form an orthonormal basis for $X$
and solve the problem
\[
\cC\phi_i=\lambda_i^2 \phi_i,\quad
i=1,2,\ldots.
\]
Furthermore, we assume that the
operator is \emph{trace-class}:
%
%e3.1 #&#
\begin{equation}
\label{eqtc} \sum_{i=1}^{\infty}
\lambda_i^2<\infty.
\end{equation}
Draws from the centred Gaussian measure
$\mu_0$ can then be made as follows.
Let $\{\xi_i\}_{i=1}^{\infty}$ denote an
independent sequence of normal random variables
with distribution $\cN(0,\lambda_i^2)$ and
consider the random function
%
%e3.2 #&#
\begin{equation}
\label{eqKL} u(x)=\sum_{i=1}^{\infty}
\xi_i \phi_i(x).
\end{equation}
This series converges in $L^2(\Omega;X)$
under the trace-class condition (\ref{eqtc}).
It is sometimes useful, both conceptually
and for purposes of implementation, to think of the unknown
function $u$ as being the infinite sequence
$\{\xi_i\}_{i=1}^{\infty}$, rather than the
function with these expansion coefficients.

We let $\cP_d$ denote projection onto the first
$d$ modes\footnote{Note that ``mode'' here, denoting an element
of a basis in a Hilbert space, differs from the ``mode'' of
a distribution.} $\{\phi_i\}_{i=1}^d$ of the Karhunen--Lo\'eve basis. Thus,
%
%e3.3 #&#
\begin{equation}
\label{eqKL2} \cP^{d_u}u(x)=\sum_{i=1}^{d_{u}}
\xi_i \phi_i(x).
\end{equation}
If the series (\ref{eqKL2}) can be summed quickly on a grid,
then this provides an efficient method for computing
exact samples from truncation of
$\mu_0$ to a finite-dimensional space.
When we refer to \emph{mesh-refine\-ment} then,
in the context of the prior, this refers to
increasing the number of terms $d_u$ used
to represent the target function $u$.

%s3.2 #&#
\subsection{Random Truncation and Sieve Priors}
\label{sssievepriors}

Non-Gaussian priors can be constructed from the Karhunen--Lo\'eve
expansion (\ref{eqKL2}) by allowing $d_u$ itself
to be a random variable supported on
$\bbN$; we let $p(i)=\bbP(d_u=i)$.
Much of the methodology in this paper
can be extended to these priors.
A draw from such a prior measure can be written as
%
%e3.4 #&#
\begin{equation}
\label{eqKL3} u(x)=\sum_{i=1}^{\infty} \bbI(i
\le d_{u})\xi_i \phi_i(x),
\end{equation}
where $\bbI(i \in E)$ is the indicator function.
We refer to this as \emph{random truncation prior}.
Functions drawn from this prior are non-Gaussian and
almost surely $C^{\infty}$. However,
expectations with respect to $d_{u}$ will be Gaussian\vadjust{\goodbreak}
and can be less regular: they are given by
the formula
%
%e3.5 #&#
\begin{equation}
\label{eqKL4} \bbE^{d_{u}} u(x)=\sum_{i=1}^{\infty}
\alpha_i \xi_i \phi_i(x),
\end{equation}
where $\alpha_i=\bbP(d_{u} \ge i)$.
As in the Gaussian case, it can be useful,
both conceptually and for purposes of implementation,
to think of the unknown
function $u$ as being the infinite vector
$ (\{\xi_i\}_{i=1}^{\infty},d_{u} )$ rather than the
function with these expansion coefficients.

Making $d_u$ a random variable has the effect of
switching on (nonzero) and off (zero) coefficients
in the expansion of the target function.
This formulation switches the
basis functions on and off in a fixed order.
Random truncation as expressed by equation (\ref{eqKL3})
is not the only variable dimension formulation. In
dimension greater than one we will employ the \emph{sieve prior}
which allows every basis function to have an individual on/off
switch. This prior relaxes the constraint imposed on the
order in which the basis functions are switched on and off
and we write
%
%e3.6 #&#
\begin{equation}
\label{eqKL5} u(x)=\sum_{i=1}^{\infty}
\chi_i \xi_i \phi_i(x),
\end{equation}
where $\{\chi_i\}_{i=1}^{\infty} \in\{0,1\}$. We
define the distribution on $\chi=\{\chi_i\}_{i=1}^{\infty}$
as follows. Let $\nu_0$ denote a reference measure
formed from considering an i.i.d. sequence
of Bernoulli random variables with success probability
one half. Then define the prior measure $\nu$ on $\chi$
to have density
\[
\frac{d\nu}{d\nu_0}(\chi) \propto\exp \Biggl(-\lambda \sum
_{i=1}^{\infty}\chi_i \Biggr),
\]
where $\lambda\in\mathbb{R}^+$.
As for the random truncation method, it is both conceptually
and practically valuable to think of the unknown function
as being the pair of random infinite vectors
$ \{ \xi_i  \}_{i=1}^\infty$ and $ \{ \chi_i
\}_{i=1}^\infty$.
Hierarchical priors, based on Gaussians but with
random switches in front of the coefficients,
are termed ``sieve priors'' in \cite{zhao2000bayesian}.
In that paper posterior consistency
questions for linear regression are also analysed
in this setting.

%s4 #&#
\section{MCMC Methods for Functions}
\label{secmcmc}

The transferable idea in this section is that
design of MCMC methods which are defined
on function spaces leads,
after discretization, to algorithms which are
robust under mesh refinement $d_{u} \to\infty$.
We demonstrate this idea for a number of algorithms,
generalizing random walk and Langevin-based
Me\-tropolis--Hastings methods, the independence
sampler, the Gibbs sampler and the HMC method;
we anticipate that
many other generalisations are possible.
In all cases the proposal exactly preserves the Gaussian
reference measure $\mu_0$ when the potential $\Phi$ is
zero and the reader may take this key idea as a design
principle for similar algorithms.

Section~\ref{ssecsetup} gives the
framework for MCMC methods on a general state
space.
In Section~\ref{ssecpropos} we state and derive
the new \emph{Crank--Nicolson} proposals, arising
from discretization of an OU process.
In Section~\ref{ssecmala} we
generalize these proposals to
the Langevin setting where steepest descent
information is incorporated: \emph{MALA proposals}.
Section~\ref{ssecind} is concerned with
\emph{Independence Samplers} which may be derived
from particular parameter choices in the
random walk algorithm.
Section~\ref{ssecrand} introduces the idea
of randomizing the choice of $\delta$ as
part of the proposal which is effective for the
random walk methods.
In Section~\ref{ssecblock} we introduce
\emph{Gibbs samplers} based on the Karhunen--Lo\'eve expansion
(\ref{eqKL}).
In Section~\ref{ssectrunc} we work with
non-Gaussian priors specified through random
truncation of the Karhunen--Lo\'eve expansion as in (\ref{eqKL3}),
showing how Gibbs samplers can again be used in this
situation.
Section~\ref{ssechmc}
briefly describes the HMC method and its generalisation
to sampling functions.

%s4.1 #&#
\subsection{Set-Up}
\label{ssecsetup}

We are interested in defining MCMC methods for measures
$\muy$ on a Hilbert space $(X, \langle\cdot,\cdot\rangle)$,
with induced norm $\|\cdot\|$,
given by (\ref{eqratz2}) where $\mu_0=\cN(0,\cC)$.
The setting we adopt is that given in \cite{Tie}
where Metropolis--Hastings methods are developed in
a general state space.
Let $q(u,\cdot)$ denote the transition
kernel on $X$ and $\eta(du,dv)$ denote the measure on $X \times X$
found by
taking $u \sim\mu$ and then $v|u \sim q(u,\cdot)$. We use $\eta
^{\perp}(u,v)$ to denote
the measure found by reversing the roles of $u$ and $v$ in the
preceding construction of $\eta$. If
$\eta^{\perp}(u,v)$ is equivalent (in the sense
of measures) to
$\eta(u,v)$, then the
Radon--Nikodym derivative $\frac{d\eta^{\perp}}{d\eta}(u,v)$ is
well-defined and we may define the
\emph{acceptance probability}
%
%e4.1 #&#
\begin{equation}
\label{eqacc}
a(u,v) = \min \biggl\{1, \frac{d\eta^\perp}{d\eta}(u,v) \biggr\}.
\end{equation}
We accept the proposed move from $u$ to $v$ with
this probability. The resulting Markov chain is $\mu$-reversible.

A key idea underlying the new variants on
random walk and Langevin-based\vadjust{\goodbreak}
Metropolis--Hastings algorithms derived
below is to use discretizations of stochastic partial
differential equations (SPDEs) which are invariant for
either the reference or the target measure.
These SPDEs have the form, for $\cL=\cC^{-1}$ the
precision operator for $\mu_0$, and $D\Phi$ the
derivative of potential $\Phi$,
%
%e4.2 #&#
\begin{equation}
\label{eqspde} \frac{du}{ds} = -\mathcal{K} \bigl(\mathcal{L}u + \gamma D
\Phi (u) \bigr) + \sqrt{2\mathcal{K}}\,\frac{db}{ds}.
\end{equation}
Here $b$ is a Brownian motion in $X$ with covariance operator the identity
and $\mathcal{K}=\mathcal{C}$ or $I$. Since
$\mathcal{K}$ is a positive operator, we may define the square-root
in the symmetric fashion, via diagonalization in the Karhunen--Lo\'eve
basis of
$\mathcal{C}$.
We refer
to it as an SPDE because in many applications $\cL$
is a differential operator. The SPDE has invariant measure
$\mu_0$ for $\gamma=0$
(when it is an infinite-dimensional OU process) and
$\mu$ for $\gamma=1$ \cite{HSWV05,HSV05,prato92}.
The target measure $\mu$
will behave like the reference measure $\mu_0$
on high frequency (rapidly oscillating) functions. Intuitively,
this is because the data, which is finite, is not
informative about the function on small scales;
mathematically, this is manifest in the absolute
continuity of $\mu$ with respect to $\mu_0$ given
by formula (\ref{eqratz2}). Thus,
discretizations of equation (\ref{eqspde}) with
either $\gamma=0$ or $\gamma=1$ form
sensible candidate proposal distributions.

The basic idea which underlies the algorithms
described here was introduced in the specific context
of conditioned diffusions with $\gamma=1$
in \cite{SVW04}, and then generalized to include the
case $\gamma=0$ in \cite{BRSV08}; furthermore, the
paper \cite{BRSV08}, although focussed on the application
to conditioned diffusions, applies to general targets of
the form (\ref{eqratz2}). The papers \cite{SVW04,BRSV08}
both include numerical results illustrating applicability
of the method to conditioned diffusion in the case $\gamma=1$,
and the paper \cite{cds11} shows application to data assimilation
with $\gamma=0$. Finally, we mention that in \cite{Neal98}
the algorithm with $\gamma=0$ is mentioned,
although the derivation does not use the SPDE
motivation that we develop here, and the concept of
a nonparametric limit is not used to motivate the
construction.

%s4.2 #&#
\subsection{Vanilla Local Proposals}
\label{ssecpropos}

The \textit{standard
random walk} proposal for $v|u$ takes the form
%
%e4.3 #&#
\begin{equation}\label{eqold}
v=u+\sqrt{2\delta\cK}\xi_0
\end{equation}
for any $\delta\in[0,\infty)$, $\xi_0 \sim\cN(0,I)$
and $\cK=I$ or $\cK=\cC$. This can be seen as a discrete
skeleton of (\ref{eqspde}) after ignoring the drift terms.
Therefore, such a proposal leads to an infinite-dimensional
version of the well-known random walk Metropolis algorithm.

The random walk proposal in finite-dimensional
problems always leads to a well-defined
algorithm and rarely encounters any reducibility problems
\cite{SmRo}. Therefore, this method can certainly
be applied for arbitrarily fine mesh size.
However, taking this approach does not lead
to a well-defined MCMC
meth\-od for \emph{functions}. This is because $\eta^{\perp}$
is singular
with respect to $\eta$ so that all proposed moves are
rejected with probability $1$.
(We prove this in Theorem~\ref{lemsing} below.)
Returning to the finite mesh case,
algorithm mixing time therefore increases to $\infty$
as $d_u \to\infty$.

To define methods with convergence properties robust to
increasing $d_u$, alternative approaches leading to
well-defined and irreducible algorithms on the Hilbert space
need to be considered. We consider two possibilities here,
both based on Crank--Nicolson approximations \cite{RM67}
of the linear part of the drift.
In particular, we consider discretization of equation
(\ref{eqspde}) with the form
%
%+ \sqrt{2\mathcal{K}}\frac{db}{ds},
%
%e4.4 #&#
\begin{eqnarray}
\label{eqcn} v &=& u - \tfrac12 \delta\mathcal{K}\mathcal{L}(u+v)\nonumber\\[-8pt]\\[-8pt]
&&{} - \delta\gamma
\mathcal{K}D\Phi(u)+\sqrt{2 \mathcal{K} \delta} \xi_0\nonumber
\end{eqnarray}
for a (spatial) white noise $\xi_0$.

%$Two modified random walks which overcome
%this difficulty, and are hence defined on
%function space, are as follows.

First consider the discretization (\ref{eqcn})
with $\gamma=0$ and $\mathcal{K}=I$.
%Multiplying through by $\cC$
Rearranging shows that the resulting
\textit{Crank--Nicolson proposal} (CN) for $v|u$ is found
by solving
%
%e4.5 #&#
\begin{equation}
\label{eqone} \bigl(I+\tfrac12\delta\cL \bigr)v= \bigl(I-\tfrac12 \delta\cL
\bigr)u + \sqrt{2\delta}\xi_0.
\end{equation}
It is this form that the proposal is best
implemented whenever the prior/reference measure
$\mu_0$ is specified via the precision operator
$\cL$ and when efficient algorithms exist for
inversion of the identity plus a multiple of $\cL$.
However, for the purposes of analysis it is also useful to
write this equation in the form
%
%e4.6 #&#
\begin{equation}
\label{eqonea} (2\cC+\delta I)v=(2\cC-\delta I)u + \sqrt{8\delta \cC} w,
\end{equation}
where $w \sim\cN(0,\cC)$,
found by applying the operator $2\cC$
to equation (\ref{eqone}).

A well-established principle in finite-dimensional
sampling algorithms advises that proposal variance should be
approximately a scalar multiple of that of the target (see,
e.g., \cite{RR01}).
The variance
in the prior,~$\mathcal{C}$, can provide a reasonable approximation,
at least as far as controlling the large $d_u$ limit is concerned.
This is because the data (or change of measure) is typically only
informative about a finite set of components in the prior model;
mathematically, the fact that the posterior has density with respect
to the prior means that it ``looks like'' the prior in the large $i$
components of the Karhunen--Lo\'eve expansion.\footnote{An interesting
research problem would be to combine the ideas in \cite{giro11},
which provide an adaptive preconditioning but are only practical
in a finite number of dimensions, with the prior-based fixed
preconditioning used here. Note that the method introduced in
\cite{giro11} reduces exactly to the preconditioning used here
in the absence of data.}

The CN algorithm violates this principle: the proposal variance
operator is proportional to $(2\mathcal{C}+\delta
I)^{-2}\cdot\mathcal {C}^2$, suggesting that algorithm
efficiency might be improved still further by obtaining a proposal
variance of~$\mathcal{C}$. In the familiar finite-dimensional case,
this can be achieved by a standard \emph{reparameterisation} argument
which has its origins in \cite {hillssmith} if not before.
%We repeat this informal argument briefly here in our context.
%Firstly note that $x=\mathcal{C}^{-1/2}u$ cons
This motivates our final local proposal in this subsection.

The \textit{preconditioned CN} proposal (pCN)
for $v|u$ is obtained from (\ref{eqcn})
with $\gamma=0$ and $\mathcal{K}=\mathcal{C}$ giving
the proposal
%
%e4.7 #&#
\begin{equation}
\label{eqtwo} (2+\delta)v=(2-\delta)u + \sqrt{8\delta} w,
\end{equation}
where, again, $w \sim\cN(0,\cC)$.
As discussed after (\ref{eqone}),
and in Section~\ref{secprior}, there
are many different ways in which the
prior Gaussian may be specified.
If the specification is via the precision $\cL$ and
if there are numerical methods for which
$(I+\zeta\cL)$ can be efficiently inverted, then
(\ref{eqone}) is a natural proposal. If,
however, sampling from $\cC$ is straightforward
(via the Karhunen--Lo\'eve expansion or directly), then
it is natural to use the proposal (\ref{eqtwo}), which
requires only that it is possible to draw
from $\mu_0$ efficiently.
For $\delta\in[0,2]$
the proposal (\ref{eqtwo}) can be written as
%
%e4.8 #&#
\begin{equation}
\label{eqsim} v = \bigl(1 - \beta^2\bigr)^{1/2}u + \beta
w,
\end{equation}
where $w \sim\mathcal{N}(0,\mathcal{C})$, and $\beta\in
[0,1]$; in fact, $\beta^2=8\delta/\break(2+\delta)^2$.
In this form we see very clearly
a simple generalisation of the finite-dimensional random walk given by
(\ref{eqold}) with $\cK=\cC$.

The numerical experiments described in Section~\ref{sseckey}
show that the pCN proposal significantly improves upon the naive
random walk method (\ref{eqold}), and similar
positive results can be obtained for the CN meth\-od.
Furthermore, for both the proposals
(\ref{eqone}) and (\ref{eqtwo}) we show in Theorem
\ref{tac} that $\eta^\perp$ and $\eta$ are equivalent
(as measures) by showing that they are both equivalent
to the same Gaussian reference measure
$\eta_0$, whilst in Theorem~\ref{lemsing}
we show that the proposal (\ref{eqold})
leads to mutually singular measures $\eta^\perp$ and $\eta$.
This theory explains the numerical observations
and motivates the importance of designing
algorithms directly on function space.

The accept--reject formula for CN and pCN
is very simple. If, for some $\rho\dvtx  X\times X \to\bbR$,
and some reference measure $\eta_0$,
%
%e4.9 #&#
\begin{eqnarray}
\label{eqstrz} \frac{d\eta}{d\eta_0}(u,v) &=& Z\exp \bigl(-\rho(u,v) \bigr),
\nonumber\\[-8pt]\\[-8pt]
\frac{d\eta^\perp}{d\eta_0}(u,v) &=& Z\exp \bigl( -\rho(v,u) \bigr),
\nonumber
\end{eqnarray}
it then follows that
%
%e4.10 #&#
\begin{equation}\label{eqthis}
\frac{d\eta^\perp}{d\eta}(u,v) = \exp \bigl(\rho(u,v)-\rho(v,u) \bigr).
\end{equation}
For both CN proposals (\ref{eqone}) and
(\ref{eqtwo}) we show in Theorem~\ref{tac} below that,
for appropriately defined $\eta_0$, we have
$\rho(u,v)=\Phi(u)$ so
that the acceptance probability is given by
%
%e4.11 #&#
\begin{equation}
\label{eqap}\quad a(u,v) = \min \bigl\{1, \exp\bigl(\Phi(u) -
\Phi(v)\bigr)\bigr \}.
\end{equation}
In this sense the CN and pCN proposals may be seen
as the \emph{natural generalisations of random walks}
to the setting where the target measure is defined
via density with respect to a Gaussian, as
in (\ref{eqratz2}). This point
of view may be understood by noting that
the accept/reject formula is defined entirely
through differences in this log density, as happens
in finite dimensions for the standard random walk,
if the density is specified with respect to the Lebesgue measure.
Similar random truncation priors are used in non-parametric inference
for drift functions in diffusion processes in~\cite{van2013reversible}.

%s4.3 #&#
\subsection{MALA Proposal Distributions}
\label{ssecmala}

The CN proposals (\ref{eqone}) and (\ref{eqtwo})
contain no information about the
potential $\Phi$ given by (\ref{eqratz2}); they contain
only information about the reference measure~$\mu_0$. Indeed,
they are derived by discretizing the SDE (\ref{eqspde})
in the case $\gamma=0$,
for which $\mu_0$ is an invariant measure.
The idea behind the Metropolis-adjusted Langevin (MALA)
proposals (see \cite{robtwe96,RC99} and the references
therein) is to discretize an equation which is invariant for the
measure $\mu$. Thus, to construct
such proposals in the function space
setting, we discretize the SPDE (\ref{eqspde}) with $\gamma=1$.
Taking $\cK=I$ and $\cK=\cC$ then gives the following two proposals.

The \textit{Crank--Nicolson Langevin proposal} (CNL)
is given by
%
%e4.12 #&#
\begin{eqnarray}
\label{three} (2\cC+\delta)v&=&(2\cC-\delta)u-2\delta\cC\mathcal{D} \Phi(u)\nonumber\\[-8pt]\\[-8pt]
&&{} +
\sqrt{8\delta\cC} w,\nonumber
\end{eqnarray}
where, as before, $w \sim\mu_0=\cN(0,\cC)$.
If we define
\begin{eqnarray*}
\rho(u,v) &=& \Phi(u) +\frac{1}{2}\bigl\langle v-u,\mathcal{D} \Phi (u)\bigr
\rangle\\
&&{}+ \frac{\delta}{4}\bigl\langle\mathcal{C}^{-1}(u+v),\mathcal{D}
\Phi (u)\bigr\rangle\\
&&{}+ \frac{\delta}{4}\bigl\| \mathcal{D} \Phi(u)\bigr\|^2,
\end{eqnarray*}
then the acceptance probability is
given by (\ref{eqacc}) and (\ref{eqthis}).
Implementation of this proposal simply requires
inversion of $(I+\zeta\cL)$, as for (\ref{eqone}).
The CNL method is the special case $\theta=\frac12$ for
the IA algorithm introduced in \cite{BRSV08}.

The \textit{preconditioned Crank--Nicolson Langevin proposal} (pCNL)
is given by
%
%e4.13 #&#
\begin{equation}
\label{four} (2+\delta)v=(2-\delta)u -2\delta\cC\mathcal{D} \Phi(u)+ \sqrt{8
\delta} w,\hspace*{-26pt}
\end{equation}
where $w$ is again a draw from $\mu_0$.
Defining
\begin{eqnarray*}
\rho(u,v) &=& \Phi(u) + \frac{1}{2}\bigl\langle v-u, \mathcal{D} \Phi (u)
\bigr\rangle\\
&&{}+ \frac{\delta}{4} \bigl\langle u+v, \mathcal{D} \Phi(u)\bigr
\rangle\\
&&{}+ \frac{\delta}{4}\bigl\|\mathcal{C}^{1/2} \mathcal{D} \Phi(u)
\bigr\|^2,
\end{eqnarray*}
the acceptance probability is given by (\ref{eqacc}) and (\ref{eqthis}).
Implementation of this proposal requires draws from
the reference measure $\mu_0$ to be made, as for (\ref{eqtwo}).
The pCNL method is the special case $\theta=\frac12$ for
the PIA algorithm introduced in \cite{BRSV08}.

%s4.4 #&#
\subsection{Independence Sampler}
\label{ssecind}

Making the choice $\delta=2$ in the pCN
proposal (\ref{eqtwo}) gives an \textit{independence sampler}.
The proposal is then simply a draw from the prior:
$v = w$.
The acceptance probability remains (\ref{eqap}).
An interesting generalisation of the independence
sampler is to take $\delta=2$ in
the MALA proposal (\ref{four}),
giving the proposal
%
%e4.14 #&#
\begin{equation}
\label{eqsim2} v = -\cC\mathcal{D} \Phi(u)+w
\end{equation}
with resulting acceptance probability given by
(\ref{eqacc}) and (\ref{eqthis}) with
\[
\rho(u,v) = \Phi(u) + \bigl\langle v, \mathcal{D} \Phi(u)\bigr\rangle+
\tfrac{1}{2}\bigl\|\mathcal{C}^{1/2} \mathcal{D} \Phi(u)\bigr\|^2.
\]

%s4.5 #&#
\subsection{Random Proposal Variance}
\label{ssecrand}

It is sometimes useful to randomise the proposal
variance $\delta$ in order to obtain better mixing.
We discuss this idea in the context of the pCN
proposal (\ref{eqtwo}).
To emphasize the dependence of the proposal kernel
on $\delta$, we denote it by $q(u,dv;\delta)$.
We show in Section~\ref{ssecthet} that
the measure $\eta_0(du,dv)=q(u,dv;\delta)\mu_0(du)$
is well-defined and symmetric in $u,v$ for every $\delta
\in[0,\infty)$. If we choose $\delta$
at random from any probability distribution $\nu$ on
$[0,\infty)$, independently from $w$,
then the resulting proposal has kernel
\[
q(u,dv)=\int_0^{\infty}q(u,dv;\delta)\nu(d\delta).
\]
Furthermore, the measure $q(u,dv)\mu_0(du)$
may be written as
\[
\int_0^{\infty} q(u,dv;\delta)\mu_0(du)
\nu(d\delta)
\]
and is hence also symmetric in $u,v$. Hence,
both the CN and pCN
proposals (\ref{eqone}) and (\ref{eqtwo}) may be generalised to
allow for $\delta$ chosen at
random independently of $u$ and $w$, according
to some measure $\nu$ on $[0,\infty)$.
The acceptance probability remains (\ref{eqap}),
as for fixed~$\delta$.

%s4.6 #&#
\subsection{Metropolis-Within-Gibbs: Blocking in Karhunen--Lo\'eve Coordinates}
\label{ssecblock}

Any function $u \in X$ can be expanded
in the\break Karhunen--Lo\'eve basis and hence written
as
%
%e4.15 #&#
\begin{equation}
\label{eqKLe} u(x)=\sum_{i=1}^{\infty}
\xi_i \phi_i(x).
\end{equation}
Thus, we may view the probability measure $\mu$ given
by (\ref{eqratz2}) as a measure on the coefficients
$u=\{\xi_i\}_{i=1}^{\infty}$. For any index set $I \subset
\bbN$ we write $\xi^{I}=\{\xi_i\}_{i \in I}$
and $\uIm=\{\xi_i\}_{i \notin I}$.
Both $\xi^{I}$ and $\uIm$ are independent
and Gaussian under
the prior $\mu_0$ with diagonal covariance
operators $\cC^{I}$ $\CIm$, respectively.
If we let $\mu_0^{I}$ denote the Gaussian $\cN(0,\cC^{I})$,
then (\ref{eqratz2}) gives
%
%e4.16 #&#
\begin{equation}
\label{eqratz3} \frac{d\muy}{d\mu_0^{I}}\bigl(\xi^{I}|\uIm\bigr) \propto
\exp \bigl(-\Phi\bigl(\xi^{I},\uIm\bigr) \bigr),
\end{equation}
where we now view $\Phi$ as a function on
the coefficients in the expansion (\ref{eqKLe}).
This formula may be used as the basis for
Metropolis-within-Gibbs samplers using blocking with respect
to a set of partitions $\{I_j\}_{j=1,\ldots, J}$
with the property $\bigcup_{j=1}^{J} I_j=\bbN$.
Because the formula is defined for functions
this will give rise to methods which are robust
under mesh refinement when implemented in practice.
We have found it useful to use the
partitions $I_j=\{j\}$ for $j=1,\ldots, J-1$
and $I_J=\{J,J+1,\ldots\}$.
On the other hand, standard Gibbs and Metropolis-within-Gibbs
samplers are based on partitioning via $I_j=\{j\}$, and do
not behave well under mesh-refinement, as we will demonstrate.

%s4.7 #&#
\subsection{Metropolis-Within-Gibbs: Random Truncation
and Sieve Priors}
\label{ssectrunc}

We will also use Metropolis-within-Gibbs to
construct sampling algorithms which alternate between
updating the
coefficients $\xi=\{\xi_i\}_{i=1}^{\infty}$ in
(\ref{eqKL3}) or (\ref{eqKL5}), and the
integer $d_u$, for (\ref{eqKL3}), or the
infinite sequence $\chi=\{\chi_i\}_{i=1}^{\infty}$
for (\ref{eqKL5}). In words, we alternate between
the coefficients in the expansion of a function
and the parameters determining which parameters
are active.

If we employ the non-Gaussian prior with draws given
by (\ref{eqKL3}), then
the negative log likelihood $\Phi$ can
be viewed as a function of $(\xi,d_u)$
and it is natural to consider
Metropolis-within-Gibbs methods which are
based on the conditional distributions for
$\xi|d_u$ and $d_u|\xi$.
Note that, under the prior, $\xi$ and $d_u$
are independent with $\xi\sim\mu_{0,\xi}:=\cN(0,\cC)$
and $d_u \sim\mu_{0,d_u}$, the latter being
supported on $\bbN$ with $p(i)=\break\bbP(d_u=i)$.
For fixed $d_u$ we have
%
%e4.17 #&#
\begin{equation}
\label{eqratz5} \frac{d\muy}{d\mu_{0,\xi}}(\xi|d_u) \propto \exp \bigl(-
\Phi(\xi,d_u) \bigr)
\end{equation}
with $\Phi(u)$ rewritten as a function of
$\xi$ and $d_u$ via the expansion (\ref{eqKL3}).
This measure can be sampled by any of the preceding
Metropolis--Hastings methods
designed in the case with Gaussian $\mu_0$.
For fixed $\xi$ we have
%
%e4.18 #&#
\begin{equation}
\label{eqratz6} \frac{d\muy}{d\mu_{0,d_u}}(d_u|\xi) \propto \exp \bigl(-
\Phi(\xi,d_u) \bigr).
\end{equation}
A natural biased random walk for $d_u|\xi$
arises by proposing moves from a random walk on $\bbN$
which satisfies detailed balance with respect to
the distribution~$p(i)$. The acceptance probability is then
\[
a(u,v) = \min \bigl\{1, \exp\bigl(\Phi(\xi,d_u) -
\Phi(\xi,d_v)\bigr)\bigr\}.
\]
Variants on this are possible and, if $p(i)$ is
monotonic decreasing, a simple random walk
proposal on the integers, with local moves
$d_u \to d_v=d_u\pm1$, is straightforward to
implement.
Of course, different proposal stencils can give improved
mixing properties, but we employ this particular
random walk for expository purposes.

If, instead of (\ref{eqKL3}),
we use the non-Gaussian sieve prior defined by
equation (\ref{eqKL5}), the prior and posterior measures may be
viewed as measures on $u =  (  \{ \xi_i  \}
_{i=1}^\infty,  \{ \chi_j  \}_{j=1}^\infty )$.
These variables may be modified as stated above via
Metropolis-within-Gibbs for sampling the conditional distributions $\xi
| \chi$ and $\chi| \xi$. If,
for example, the proposal for $\chi| \xi$ is reversible
with respect to the prior on $\xi$, then the acceptance probability
for this move is given by
\[
a ( u,v ) = \min \bigl\{1, \exp \bigl( \Phi ( \xi _u,
\chi_u ) - \Phi ( \xi_v,\chi_v ) \bigr)
\bigr\}.
\]
In Section~\ref{ssecgeo2} we implement a slightly
different proposal in which, with probability $\frac12$, a
nonactive mode is switched on with the remaining
probability\vspace*{1pt} an active mode is switched off. If
we define $N_{\mathrm{on}} = \sum_{i=1}^N \chi_i$, then
the probability of moving from $\xi_u$ to a state
$\xi_v$ in which an extra mode is switched on is
\begin{eqnarray*}
a ( u,v )
&=& \min \biggl\{1, \exp \biggl( \Phi ( \xi _u,
\chi_u ) - \Phi ( \xi_v,\chi_v )\\[-1.5pt]
&&\hspace*{103.2pt}{} +
\frac
{N-N_{\mathrm{on}}}{N_{\mathrm{on}}} \biggr) \biggr\}.
\end{eqnarray*}
Similarly, the probability of moving to a situation
in which a mode is switched off is
\begin{eqnarray*}
a ( u,v ) &=& \min \biggl\{1, \exp \biggl( \Phi ( \xi _u,
\chi_u ) - \Phi ( \xi_v,\chi_v ) \\[-1.5pt]
&&\hspace*{102.3pt}{}+
\frac
{N_{\mathrm{on}}}{N-N_{\mathrm{on}}} \biggr) \biggr\}.
\end{eqnarray*}

%s4.8 #&#
\subsection{Hybrid Monte Carlo Methods}
\label{ssechmc}

The algorithms discussed above have been based
on proposals which can be motivated through
discretization of an SPDE which is invariant
for either the prior measure $\mu_0$ or for the
posterior $\mu$ itself. HMC methods are based
on a different idea, which is to consider
a Hamiltonian flow in a state space found from
introducing extra ``momentum'' or ``velocity'' variables
to complement the variable $u$ in (\ref{eqratz2}).
If the momentum/velocity is chosen randomly
from an appropriate Gaussian distribution at regular
intervals, then the resulting Markov chain in $u$
is invariant under $\mu$. Discretizing the flow,
and adding an accept/reject step, results in
a method which remains invariant for $\mu$ \cite{duna87}.
These methods can break random-walk type behaviour
of methods based on local proposal \cite{neal96,neal2010}.
It is hence of interest to generalise these methods
to the function sampling setting dictated by (\ref{eqratz2})
and this is undertaken in \cite{besk10b}.
The key novel idea required to design this algorithm
is the development of a new integrator for the Hamiltonian
flow underlying the method; this integrator is exact
in the Gaussian case $\Phi\equiv0$, on function space,
and for this reason
behaves well for nonparametric where $d_u$ may be
arbitrarily large infinite dimensions.

%s5 #&#
\section{Computational Illustrations}

This section contains numerical experiments
designed to illustrate various properties of the
sampling algorithms overviewed in this paper.
We employ the examples introduced in Section~\ref{seccom}.

%s5.1 #&#
\subsection{Density Estimation}
\label{ssecden2}

Section~\ref{sseckey} shows an example which
illustrates the advantage of using the function-space
algorithms highlighted in this paper in comparison
with standard techniques; there we compared pCN
with a standard random walk.
The first goal of the experiments in this subsection
is to further illustrate the advantage of the
function-space algorithms over standard algorithms.
Specifically, we compare the Metropolis-within-Gibbs
method from Section~\ref{ssecblock}, based
on the partition $I_j=\{j\}$ and labelled MwG here, with
the pCN sampler from Section~\ref{ssecpropos}.
The second goal is to study the effect of prior modelling
on algorithmic performance; to do this, we study
a third algorithm, RTM-pCN, based on sampling the randomly
truncated Gaussian prior (\ref{eqKL3})
using the Gibbs method from Section~\ref{ssectrunc},
with the pCN sampler for the coefficient update.

%s5.1.1 #&#
\subsubsection{Target distribution}

We will use the true density
\begin{eqnarray*}
\rho&\propto&\mathcal{N} ( -3, 1 ) \bbI \bigl(x \in ( -\ell, +\ell )
\bigr) \\
&&{}+ \mathcal{N} ( +3, 1 ) \bbI \bigl( x \in ( -\ell, +\ell ) \bigr),
\end{eqnarray*}
where $\ell=10$. [Recall that $\bbI(\cdot)$ denotes
the indicator function of a set.] This density corresponds
approximately to a situation where there is a $50/50$ chance
of being in one of the two Gaussians.
This one-dimensional multi-modal density
is sufficient to expose the advantages of
the function spaces samplers pCN and RTM-pCN over MwG.

%
%f2 #&#
\begin{figure*}

\includegraphics{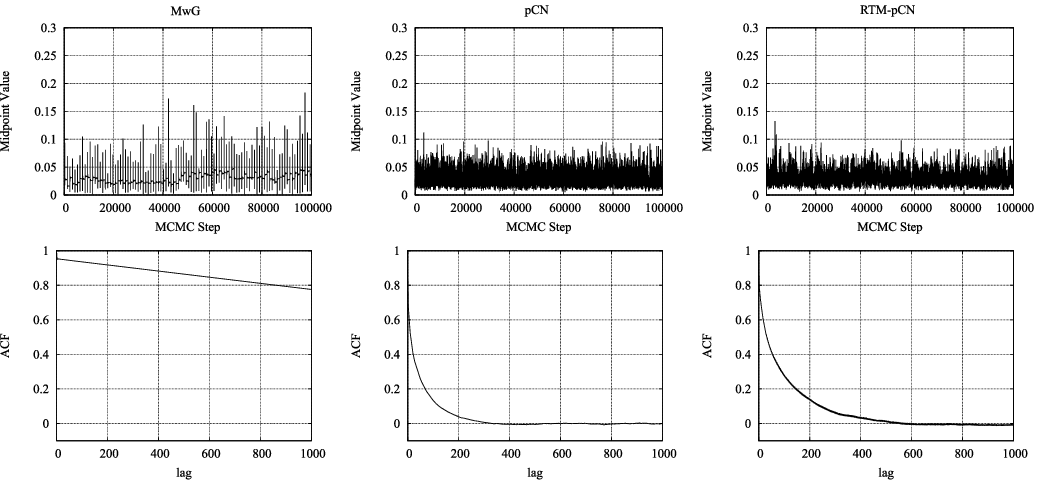}

\caption{Trace and autocorrelation plots for sampling posterior measure
with true density $\rho$ using MwG, pCN and RTM-pCN methods.}
\label{cfbimodalgauss}
\end{figure*}
%

%s5.1.2 #&#
\subsubsection{Prior}

We will make comparisons between the three algorithms regarding their
computational performance, via various graphical and numerical
measures. In all cases it is important that the reader appreciates
that the comparison between MwG and pCN corresponds to sampling from
the same posterior, since they use the same prior, and that all
comparisons between RTM-pCN and other methods also quantify the effect of
prior modelling as well as algorithm.

Two priors are used for this experiment: the Gaussian
prior given by (\ref{eqKL}) and the randomly truncated
Gaussian given by (\ref{eqKL3}).
We apply the MwG and pCN schemes in the former case,
and the RTM-pCN scheme for the latter.
The prior uses the same Gaussian covariance structure for
the independent~$\xi$, namely,
$\xi_i \sim\mathcal{N}  ( 0, \lambda_i^2  )$,
where $\lambda_i \propto i^{-2}$.
Note that the eigenvalues are summable, as required
for draws from the Gaussian measure to be square integrable
and to be continuous.
The prior for the number of active terms $d_u$
is an exponential distribution with rate $\lambda=0.01$.

%s5.1.3 #&#
\subsubsection{Numerical implementation}

In order to facilitate a fair comparison, we tuned the
value of $\delta$ in the pCN and RTM-pCN proposals to
obtain an average acceptance probability of around
$0.234$, requiring, in both cases, $\delta\approx0.27$.
(For RTM-pCN the average acceptance probability
refers only to moves in $ \{ \xi \}_{i=1}^\infty$
and not in $d_u$.)
We note that with the value $\delta=2$ we obtain
the independence sampler for pCN; however, this
sampler only accepted $12$ proposals out of $10^6$ MCMC steps,
indicating the importance of tuning $\delta$ correctly.
For MwG there is no tunable parameter, and we obtain
an acceptance of around $0.99$.

% For the second experiment, involving $\rho_2$, a
% similar procedure of tuning the step size to attain an average
%acceptance of $0.234$ was undertaken.
% On this heavily frequency dependent problem the RTM-pCN
% permitted a slightly larger step size, for the same
% acceptance probability, than pCN, as shown in Table~\ref{table}.
% It is shown below that this larger step size results in smaller
% correlations between MCMC samples.

% \medskip
% \begin{table}
% \begin{center}
% \begin{tabular}{ | c | c | c |}
% \hline
% Algorithm & $\delta$ & $\mathbb{E} \left[ \alpha\right]$ \\
% \hline
% MwG & NA & 0.777 \\
% pCN & 0.065 & 0.241 \\
% RTM-pCN & 0.07 & 0.224 \\
% \hline
% \end{tabular}
% \caption{Average Acceptance Probabilities and Proposal Variance for
% Target $\rho_2$}
% \label{table}
% \end{center}
% \end{table}
%
%t1 #&#
\begin{table}
\tablewidth=150pt
\caption{Approximate integrated autocorrelation times for target $\rho$}
\label{tabdeniactrho1}
\begin{tabular*}{\tablewidth}{@{\extracolsep{\fill}}l d{3.1}@{}}
\hline
\textbf{Algorithm} & \multicolumn{1}{c@{}}{\textbf{IACT}} \\
\hline
MwG & 894 \\
pCN & 73.2 \\
RTM-pCN & 143 \\
\hline
\end{tabular*}
\end{table}

%s5.1.4 #&#
\subsubsection{Numerical results}

In order to compare the performance of pCN, MwG and RTM-pCN,
we show, in Figure~\ref{cfbimodalgauss} and Table~\ref{tabdeniactrho1},
trace plots, correlation functions and integrated auto-correlation times
(the latter are notoriously
difficult to compute accurately \cite{sokal} and displayed
numbers to three significant figures should only be treated
as indicative).
The autocorrelation function decays for ergodic Markov chains,
and its integral determines the asymptotic variance of
sample path averages.
The integrated autocorrelation time
is used, via this asymptotic variance,
to determine the number of steps required to
determine an independent sample from the MCMC method.
%These quantities are calculated using the values $x \left( 0 \right)$.
The figures and integrated autocorrelation times clearly
show that the pCN and RTM-pCN outperform MwG by an order
of magnitude. This reflects the fact that pCN and RTM-pCN
are function space samplers, designed to mix independently
of the mesh-size. In contrast, the MwG method is heavily
mesh-dependent, since updates are made one Fourier
component at a time.

%
%t2 #&#
\begin{table}
\caption{Comparison of computational timings for target $\rho$}
\label{tabdenruntimesrho1}
\begin{tabular*}{\tablewidth}{@{\extracolsep{\fill}}lcd{1.4}@{}}
\hline
&& \multicolumn{1}{c@{}}{\textbf{Time to draw an}}\\
\textbf{Algorithm} & \textbf{Time for} $\bolds{10^6}$ \textbf{steps (s)}
& \multicolumn{1}{c@{}}{\textbf{indep sample
(s)}} \\
\hline
MwG & $262$ & 0.234 \\
pCN & $451$ & 0.0331 \\
RTM-pCN & $278$ & 0.0398 \\
\hline
\end{tabular*}
\end{table}

Finally, we comment on the effect of the
different priors. The asymptotic
variance for the RTM-pCN is approximately double that of pCN.
However, RTM-pCN can have a reduced runtime, per unit error,
when compared with pCN, as Table~\ref{tabdenruntimesrho1}
shows. This improvement of RTM-pCN over pCN is primarily caused by
the reduction in the number of random number generations
due to the adaptive size of the basis in which the unknown density
is represented.
\subsection{Data Assimilation in Fluid Mechanics}
\label{ssecda2}

We now proceed to a more complex problem and describe numerical
results which demonstrate that the function space samplers
successfully
sample nontrivial problems arising in applications. We study both the
Eulerian and Lagrangian data assimilation problems from Section~\ref{ssecda}, for the Stokes flow forward model $\gamma=0$.
It has been demonstrated in \cite{SC10,cds11} that the pCN can successfully
sample from the posterior distribution for such problems.
In this subsection we
will illustrate three features of such methods:
convergence of the
algorithm from different starting states, convergence with
different proposal step sizes, and behaviour
with random distributions for the proposal step size,
as discussed in Section~\ref{ssecrand}.

%s5.2.1 #&#
\subsubsection{Target distributions}
In this application we aim to characterize the posterior distribution
on the initial condition of the two-dimensional velocity field
$u_0$ for Stokes flow [equation (\ref{NS}) with $\gamma=0$],
given a set of either Eulerian (\ref{eedata}) or
Lagrangian (\ref{eldata}) observations.
In both cases, the posterior is of the form
(\ref{eqratz2}) with $\Phi(u) = \frac{1}{2} \|\mathcal{G}(u) -
y\|^2_\Gamma$, with $\mathcal{G}$ a nonlinear mapping
taking $u$ to the observation space.
We choose the observational noise covariance
to be $\Gamma= \sigma^2I$ with $\sigma= 10^{-2}$.

%s5.2.2 #&#
\subsubsection{Prior}
We let $A$ be the Stokes operator defined by writing (\ref{NS}) as
$dv/dt+Av=0, v(0)=u$ in the case $\gamma=0$ and $\psi=0$. Thus, $A$ is
$\nu$ times the negative Laplacian,\vadjust{\goodbreak} restricted to a divergence free
space; we also work on the space of functions whose spatial average is
zero and then $A$ is invertible. For the numerics that follow, we set
$\nu= 0.05$. It is important to note that, in the periodic setting
adopted here, $A$ is diagonalized in the basis of divergence free
Fourier series. Thus, fractional powers of $A$ are easily calculated.
The prior measure is then chosen as
%
%e5.1 #&#
\begin{equation}
\label{eqprior} \mu_0=\mathcal{N}\bigl(0,\delta A^{-\alpha}
\bigr),
\end{equation}
in both the Eulerian and Lagrangian data scenarios. We
require $\alpha> 1$ to ensure that the
eigenvalues of the covariance are summable
(a necessary and sufficient condition
for draws from the prior, and hence the
posterior, to be continuous functions, almost surely).
In the numerics that follow, the parameters of
the prior were chosen
to be $\delta= 400$ and $\alpha= 2$.

%s5.2.3 #&#
\subsubsection{Numerical implementation}
The figures that follow in this section are taken from what are termed
\emph{identical twin} experiments in the data assimilation community:
the same approximation of the model described above to simulate the
data is also used for evaluation of $\Phi$ in the statistical
algorithm in the calculation of the likelihood of $u_0$ given the
data, with the same assumed covariance structure of the observational
noise as was used to simulate the data.

Since the domain is the two-dimensional torus, the evolution of the
velocity field can be solved exactly for a truncated Fourier series,
and in the numerics that follow we truncate this to $100$ unknowns, as
we have found the results to be robust to further refinement. In the
case of the Lagrangian data, we integrate the trajectories
(\ref{eqlag}) using an Euler scheme with time step $\Delta t =
0.01$. In each case we will give the values of $N$ (number of spatial
observations, or particles) and $M$ (number of temporal observations)
that were used. The observation stations (Eulerian data) or initial
positions of the particles (Lagrangian data) are evenly spaced on a
grid. The $M$ observation times are evenly spaced, with the final
observation time given by $T_M=1$ for Lagrangian observations and
$T_M=0.1$ for Eulerian. The true initial condition $u$ is chosen
randomly from the prior distribution.

%s5.2.4 #&#
\subsubsection{Convergence from different initial states}

We consider a posterior distribution found from data comprised of
$900$ Lagrangian tracers observed at $100$ evenly spaced times on
$[0,1]$. The data volume is high and a form of posterior consistency
is observed for low Fourier modes, meaning that the posterior is
approximately a Dirac mass at the truth. Observations were made of\vadjust{\goodbreak}
%
%f3 #&#
\begin{figure}

\includegraphics{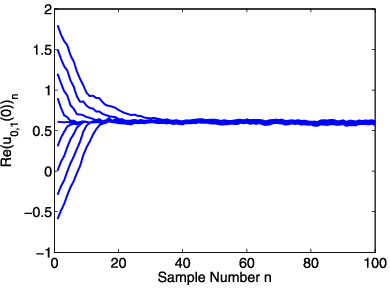}

\caption{Convergence of value of one Fourier mode of the initial
condition $u_0$ in the pCN Markov chains with different initial states,
with Lagrangian data.} \label{LagSDP}
\end{figure}
each of these tracers up to a final time $T=1$. Figure~\ref{LagSDP}
shows traces of the value of one particular Fourier mode\footnote{The
real part of the coefficient of the Fourier mode with wave number 0
in the $x$-direction and wave number $1$ in the $y$-direction.} of the
true initial conditions. Different starting values are used for pCN
and all converge to the same distribution. The proposal variance
$\beta$ was chosen in order to give an average acceptance probability
of approximately $25\%$.

%s5.2.5 #&#
\subsubsection{Convergence with different $\beta$}

\begin{figure}

\includegraphics{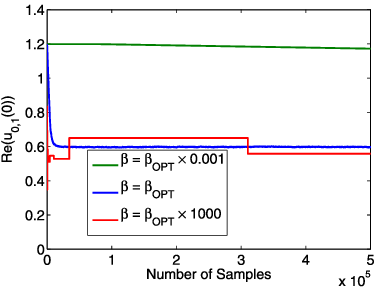}

\caption{Convergence of one of the Fourier modes of the initial
condition in the pCN Markov chains with different proposal variances,
with Eulerian data.} \label{diffBetaConv}
\end{figure}

Here we study the effect of varying the proposal variance. Eulerian
data is used with $900$ observations in space and $100$ observation
times on $[0,1]$. Figure~\ref{diffBetaConv} shows the different rates
%
%f4 #&#
%
%
%f5 #&#
\begin{figure*}

\includegraphics{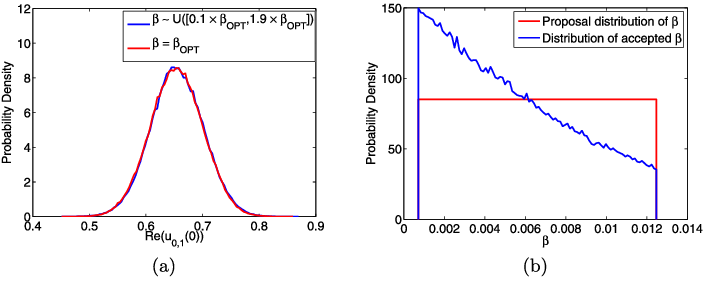}

\caption{Eulerian data assimilation example. \textup{(a)} Empirical marginal
distributions estimated using the pCN with and without random $\beta$.
\textup{(b)}~Plots of the proposal distribution for $\beta$ and the distribution
of values for which the pCN proposal was accepted.} \label{RandBetaR01}
\end{figure*}
of convergence of the algorithm with different values of $\beta$, in
the same Fourier mode coefficient as used in Figure~\ref{LagSDP}. The
value labelled $\beta_{\mathrm{opt}}$ here is chosen to give an acceptance
rate of approximately $50\%$. This value of $\beta$ is obtained by
using an adaptive burn-in, in which\vadjust{\goodbreak} the acceptance probability is
estimated over short bursts and the step size $\beta$ adapted
accordingly. With $\beta$ too small, the algorithm accepts proposed
states often, but these changes in state are too small, so the
algorithm does not explore the state space efficiently. In contrast,
with $\beta$ too big, larger jumps are proposed, but are often
rejected since the proposal often has small probability density and
so are often rejected. Figure~\ref{diffBetaConv} shows examples of
both of these, as well as a more efficient choice $\beta_{\mathrm{opt}}$.

%s5.2.6 #&#
\subsubsection{\texorpdfstring{Convergence with random $\beta$}{Convergence with random beta}}

Here we illustrate the possibility of using a random proposal variance
$\beta$, as introduced in Section~\ref{ssecrand} [expressed in terms
of $\delta$ and (\ref{eqtwo}) rather than $\beta$ and (\ref{eqsim})].
Such methods have the potential advantage of including the possibility
of large and small steps in the proposal. In this example we use
Eulerian data once again, this time with only $9$ observation stations,
with only one observation time at $T=0.1$. Two instances of the sampler
were run with the same data, one with a static value of $\beta=
\beta_{\mathrm{opt}}$ and one with $\beta\sim U([0.1 \times\bopt, 1.9
\times\bopt])$. The marginal distributions for both Markov chains are
shown in Figure~\ref{RandBetaR01}(a), and are very close indeed,
verifying that randomness in the proposal variance scale gives rise to
(empirically) ergodic Markov chains. Figure~\ref{RandBetaR01}(b) shows
the distribution of the $\beta$ for which the proposed state was
accepted. As expected, the initial uniform distribution is skewed, as
proposals with smaller jumps are more likely to be accepted.

The convergence of the method with these two choices for $\beta$ were
roughly comparable in this simple experiment. However, it is
of course conceivable that when attempting to explore multimodal posterior
distributions it may be advantageous to have a mix of both large
proposal steps, which may allow large leaps between different areas of
high probability density, and smaller proposal steps in order to
explore a more localised region.

%s5.3 #&#
\subsection{Subsurface Geophysics}
\label{ssecgeo2}

The purpose of this section is twofold: we demonstrate
another nontrivial application where function space
sampling is potentially useful and we demonstrate the use
of sieve priors in this context. Key to understanding
what follows in this problem is to appreciate that,
for the data volume we employ, the posterior distribution
can be very diffuse and expensive to explore unless
severe prior modelling is imposed, meaning that
the prior is heavily weighted to solutions with only
a small number of active Fourier modes, at low wave numbers.\vadjust{\goodbreak}
This is because
the homogenizing property of the elliptic PDE means that a
whole range of different length-scale solutions can explain
the same data. To combat this, we choose
very restrictive priors,
either through the form of Gaussian covariance or through
the sieve mechanism, which favour a small number of active
Fourier modes.

%s5.3.1 #&#
\subsubsection{Target distributions}

\begin{figure}

\includegraphics{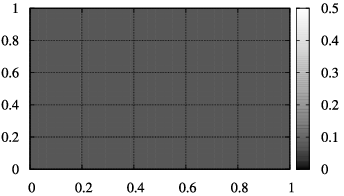}

\caption{True permeability function used to create target
distributions in subsurface geophysics application.}
\label{geofigkappatrue99}
\end{figure}

\begin{figure}

\includegraphics{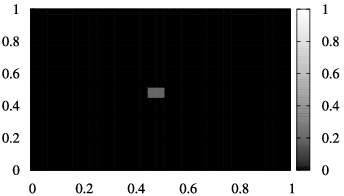}

\caption{Measurement locations for subsurface experiment.}
\label{geofigmeasurelocs}
\end{figure}

We consider equation (\ref{eqhead}) in the case
$D=[0,1]^2$.
Recall that the objective in this problem is to recover
the permeability $\kappa=\exp ( u  )$.
The sampling algorithms discussed here are applied to
the log permeability $u$. The ``true'' permeability for which we test
the algorithms is shown in Figure~\ref{geofigkappatrue99} and is
given by
%
%e5.2 #&#
\begin{equation}
\label{geoeqkappaconst} \kappa ( x ) =
\exp \bigl( u_1 ( x ) \bigr) = \tfrac{1}{10}.
\end{equation}
The pressure measurement data is
$y_j= p  ( x_j  )+\sigma\eta_j$
with the $\eta_j$ i.i.d. standard unit Gaussians,
with the measurement location
shown in Figure~\ref{geofigmeasurelocs}.

%s5.3.2 #&#
\subsubsection{Prior}
The priors will either be Gaussian or a sieve prior
based on a Gaussian. In both cases the Gaussian
structure is defined via a Karhunen--Lo\'eve expansion of the form
%
%e5.3 #&#
\begin{eqnarray}
\label{geoeqprior} u ( x ) &=& \zeta_{0,0}
\varphi^{  ( 0,0  ) }\nonumber\\[-8pt]\\[-8pt]
&&{}+ \sum_{  ( p,q ) \in\bbZ^2 \diagup \{0,0 \}}
\frac{ \zeta_{p,q} \varphi^{  ( p,q  ) } }{  ( p^2 +
q^2  )^\alpha},\nonumber
\end{eqnarray}
where $\varphi^{  ( p,q  ) }$ are two-dimensional
Fourier basis functions and the $\zeta_{p,q}$ are
independent random variables with distribution
$\zeta_{p,q} \sim\mathcal{N}  ( 0,1  )$
and $a \in\mathbb{R}$. To ensure that the eigenvalues
of the prior covariance operator are summable
(a necessary and sufficient condition
for draws from it to be continuous functions, almost surely),
we require that $\alpha> 1$.
For target defined via $\kappa$ we
take $\alpha= 1.001$.\vadjust{\goodbreak}

%
%f6 #&#
%

%f7 #&#

For the Gaussian prior we employ MwG and pCN schemes, and
we employ the pCN-based Gibbs sampler from
Section~\ref{ssectrunc} for the sieve prior; we refer to
this latter algorithm as Sieve-pCN. As in Section~\ref{ssecden2}, it is important
that the reader appreciates that the comparison between MwG and pCN
corresponds to sampling from the same posterior, since they use the
same prior, but that all comparisons between Sieve-pCN and other methods
also quantify the effect of prior modelling as well as algorithm.

%s5.3.3 #&#
\subsubsection{Numerical implementation\hspace*{-1pt}}

The forward mod\-el is evaluated by solving equation (\ref{eqhead})
on the two-dimensional domain $D =  [ 0,1  ]^2$ using
a finite difference method with mesh of size $J \times J$.
This results in a $J^2 \times J^2$ banded matrix with
bandwidth $J$ which may be solved, using a banded matrix solver,
in $\mathcal{O}  ( J^4  )$ floating point operations
(see page $171$ \cite{iserles04}). As drawing a sample is a
$\mathcal{O}  ( J^4  )$ operation, the grid sizes
used within these experiments was kept deliberately low:
for target defined via $\kappa$ we take $J=64$.
This allowed
a sample to be drawn in less than $100$ ms and therefore $10^6$
samples to be drawn in around a day.
We used $1$ measurement point, as shown in
Figure~\ref{geofigmeasurelocs}.

%s5.3.4 #&#
\subsubsection{Numerical results}

\begin{figure}

\includegraphics{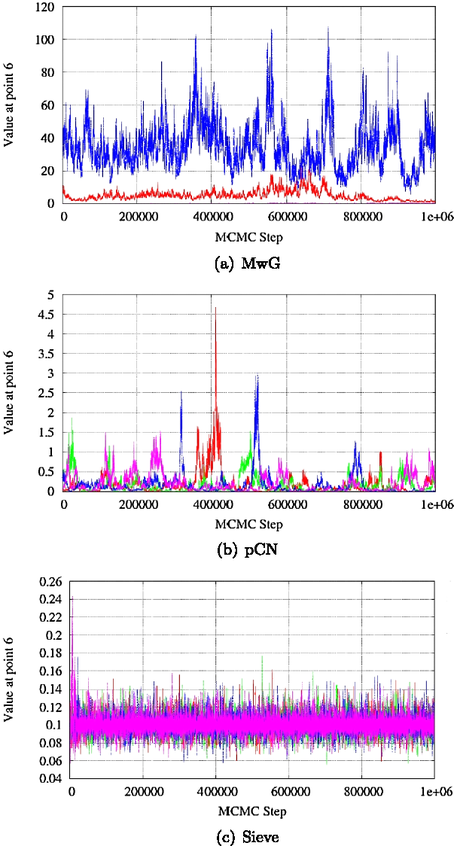}

\caption{Trace plots for the subsurface geophysics
application, using 1 measurement. The MwG, pCN and Sieve-pCN
algorithms are compared. Different colours correspond to identical
MCMC simulations with different random number generator seeds.}
\label{geofigbroadpriorconvergetracesplots}
\end{figure}

Since $\alpha=1.001$, the ei\-gen\-values of the prior covariance
are only just summa\-ble, meaning that many Fourier modes will
be active in the prior.
Figure~\ref{geofigbroadpriorconvergetracesplots} shows
%
%f8 #&#
trace plots obtained through application of the MwG and
pCN methods to the Gaussian prior and a pCN-based Gibbs
sampler for the sieve prior, denoted Sieve-pCN.
The proposal variance for pCN and Sieve-pCN
was selected to ensure an average acceptance of around $0.234$.
Four different seeds are used.
It is clear from these plots that only the
MCMC chain generated
by the sieve prior/algorithm
combination converges in the available
computational time. The other algorithms fail to converge
under these test conditions.
This demonstrates the importance of prior modelling
assumptions for these under-determined inverse problems
with multiple solutions.

\subsection{Image Registration}
\label{ssecim2}

In this subsection we consider the image registration
problem from Section~\ref{ssecim}. Our primary purpose
is to illustrate the idea that, in the function space
setting, it is possible to extend the prior modelling
to include an unknown observational precision and to
use conjugate Gamma priors for this parameter.\looseness=1

%s5.4.1 #&#
\subsubsection{Target distribution}

We study the setup from Section~\ref{ssecim},
with data generated from a noisily observed truth
$u=(p,\eta)$ which corresponds to a smooth closed
curve. We make $J$ noisy observations of the
curve where, as will be seen below, we consider
the cases $J=10,20,50,100,200,500$ and $1000$. The
noise used to generate the data
is an uncorrelated mean zero Gaussian at each
location with variance $\sigma_{\mathrm{true}}^2=0.01$. We
will study the case where the noise variance
$\sigma^2$ is itself considered unknown,
introducing a prior on $\tau=\sigma^{-2}$.
We then use MCMC to study the
posterior distribution on $(u,\tau)$, and hence
on $(u,\sigma^2)$.\looseness=1

%f9 #&#
\begin{figure*}

\includegraphics{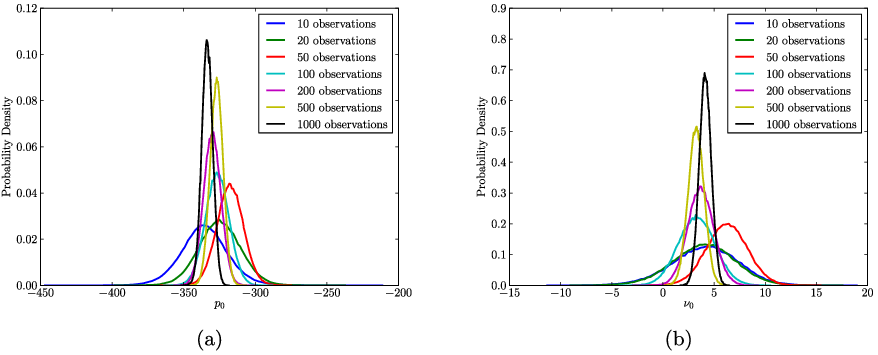}

\caption{Convergence of the lowest wave number Fourier modes in \textup{(a)} the
initial momentum $P_0$, and \textup{(b)} the reparameterisation function $\nu$,
as the number of observations is increased, using the pCN.}\label{pnu}
\end{figure*}

%s5.4.2 #&#
\subsubsection{Prior}

The priors on the initial momentum and reparameterisation
are taken as
%
%e5.4 #&#
\begin{eqnarray}
\label{pnupriors} \mu_{p}(p) &=& \mathcal{N}\bigl(0,
\delta_1\cH^{-\alpha_1}\bigr),
\nonumber\\[-8pt]\\[-8pt]
\mu_{\nu}(\nu) &=& \mathcal{N}\bigl(0,\delta_2
\cH^{-\alpha_2}\bigr),
\nonumber
\end{eqnarray}
where $\alpha_1 = 0.55$, $\alpha_2 = 1.55$, $\delta_1 = 30$ and
$\delta_2 = 5\cdot10^{-2}$. Here $\cH= (I - \Delta)$ denotes the
Helmholtz operator in one dimension and, hence, the
chosen values of $\alpha_i$ ensure that the eigenvalues of the
prior covariance operators are summable.
As a consequence, draws from the prior are
continuous, almost surely.
The prior for $\tau$ is defined as
%
%e5.5 #&#
\begin{equation}
\mu_{\tau} = \operatorname{Gamma}(\alpha_\sigma,\beta_\sigma),
\end{equation}
noting that this leads to a conjugate posterior on this
variable, since the observational noise is Gaussian.
In the numerics that follow, we set
$\alpha_\sigma= \beta_\sigma= 0.0001$.

%s5.4.3 #&#
\subsubsection{Numerical implementation}
In each experiment the data is produced using the same template shape
$\Gd$, with parameterization given by
%
%e5.6 #&#
\begin{eqnarray}
\label{template} \qd(s) = \bigl(\cos(s) + \pi,\sin(s) + \pi \bigr),\nonumber\\[-8pt]\\[-8pt]
\eqntext{s\in [0,2
\pi).}
\end{eqnarray}
In the following numerics, the observed shape is chosen by first
sampling an instance of $p$ and $\nu$ from their respective prior
distributions and using the numerical approximation
of the forward model to give us the parameterization of the
target shape. The $N$ observational points
$\{s_i\}_{i=1}^N$ are then picked by evenly spacing them out over the
interval $[0,1)$, so that $s_i = (i-1)/N$.

%s5.4.4 #&#
\subsubsection{Finding observational noise hyperparameter}

We implement an MCMC method to sample from the joint
distribution of $(u,\tau)$, where (recall) $\tau=\sigma^{-2}$
is the inverse observational precision. When sampling
$u$ we employ the pCN method. In this context
it is possible to either: (i) implement a Metropolis-within-Gibbs
sampler, alternating between use of pCN to sample $u|\tau$
and using explicit sampling from the Gamma distribution
for $\tau|u$; or (ii) marginalize out $\tau$ and sample
directly from the marginal distribution for $u$, generating
samples from $\tau$ separately; we adopt the second
approach.

We show that, by taking data sets with an increasing number
of observations $N$, the true values of the functions $u$
and the precision parameter $\tau$ can both be
recovered: a form of posterior consistency.

\begin{figure}

\includegraphics[scale=0.99]{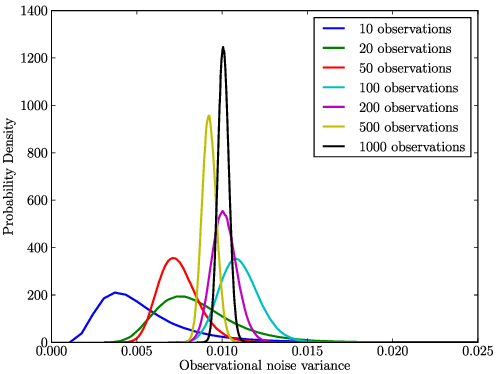}

\caption{Convergence of the posterior distribution on the value of the
noise variance $\sigma^2 I$, as the number of observations is
increased, sampled using the pCN.}
\label{p1NTsig}
\end{figure}

This is demonstrated in Figure~\ref{pnu}, for the
posterior distribution on a low wave number Fourier coefficient
in the expansion of the initial momentum $p$ and the
reparameterisation $\eta$.
Figure~\ref{p1NTsig} shows the posterior distribution on the
%
%f10 #&#
value of the observational variance~$\sigma^2$; recall
that the true value is $0.01$.
The posterior distribution becomes
increasingly peaked close to this value as $N$ increases.

%s5.5 #&#
\subsection{Conditioned Diffusions}
\label{sseccond2}

Numerical experiments which employ function\break space samplers to study
problems arising in conditioned diffusions have been published in a
number of articles. The paper \cite{BRSV08} introduced the idea of
function space samplers in this context and demonstrated the advantage
of the CNL method (\ref{three}) over the standard Langevin algorithm
for bridge diffusions; in the notation of that paper, the IA method
with $\theta=\frac12$ is our CNL method. Figures analogous to Figure
\ref{fig1}(a) and (b) are shown. The article \cite{HSV09} demonstrates
the effectiveness of the CNL method, for smoothing problems arising in
signal processing, and figures analogous to Figure~\ref{fig1}(a) and
(b) are again shown. The paper \cite{besk10b} contains numerical
experiments showing comparison of the function-space HMC method from
Section~\ref{ssechmc} with the CNL variant of the MALA method from
Section~\ref{ssecmala}, for a bridge diffusion problem; the
function-space HMC method is superior in that context, demonstrating
the power of methods which break random-walk type behaviour of local
proposals.

%s6 #&#
\section{Theoretical Analysis}
\label{secInvP}

The numerical experiments in this paper demonstrate
that the function-space algorithms of Crank--Nicolson
type behave well on a range of nontrivial examples.
In this section we describe some theoretical analysis
which adds weight to the
choice of Crank--Nicolson discretizations which
underlie these algorithms. We also show
that the acceptance probability resulting from
these proposals behaves as in finite dimensions:
in particular, that it is continuous as the scale factor
$\delta$ for the proposal variance tends to zero.
And finally we summarize briefly the theory available
in the literature which relates to the function-space
viewpoint that we highlight in this paper.
We assume throughout that $\Phi$ satisfies the following
assumptions:

%as6.1 #&#
\begin{asp}
\label{ass1}
The function $\Phi\dvtx X\to\bbR$ satisfies the following:
\begin{enumerate}
\item there exists $p>0,K>0$
such that, for all $u \in X$
\[
0 \le\Phi(u;y) \le K \bigl(1+\|u\|^p\bigr);
\]
\item for every $r>0$ there is $K(r)>0$ such that, for
all $u,v \in X$
with $\max\{\|u\|,\|v\|\}<r$,
\[
\bigl|\Phi(u)-\Phi(v)\bigr| \le K(r)\|u-v\|.
\]
\end{enumerate}
\end{asp}

These assumptions arise naturally in many Bayes\-ian
inverse problems where the data is finite
dimensional \cite{Stuart10}.
Both the data assimilation inverse problems from
Section~\ref{ssecda} are shown to satisfy
Assumptions~\ref{ass1},
for appropriate choice of $X$ in \cite{cdrs08}
(Navier--Stokes) and \cite{Stuart10} (Stokes).
The groundwater flow inverse\vadjust{\goodbreak} problem from Section~\ref{ssecgeo} is shown
to satisfy these assumptions in \cite{ClDS10},
again for approximate choice of $X$.
It is shown in \cite{CC10} that
Assumptions~\ref{ass1} are satisfied for
the image registration problem of Section~\ref{ssecim},
again for appropriate choice of $X$.
A~wide range of conditioned diffusions
satisfy Assumptions
\ref{ass1}; see \cite{HSV10}.
The density estimation problem from Section~\ref{ssecden}
satisfies the second item from Assumptions~\ref{ass1}, but
not the first.

%s6.1 #&#
\subsection{Why the Crank--Nicolson Choice?}
\label{ssecthet}

In order to explain this choice, we
consider a one-parameter ($\theta$)
family of discretizations
of equation (\ref{eqspde}),
which reduces to the discretization (\ref{eqcn}) when
$\theta=\frac12$. This family is
%
%e6.1 #&#
\begin{eqnarray}
\label{eqprop0}\label{eqyarra}\qquad
v&=&u-\delta \mathcal{K} \bigl((1-\theta)\mathcal{L}u+\theta
\mathcal{L}v \bigr) -\delta\gamma\mathcal{K}D\Phi(u)\nonumber\\[-8pt]\\[-8pt]
&&{}+ \sqrt{2\delta\mathcal{K}}
\xi_0,\nonumber
\end{eqnarray}
where $\xi_0 \sim\mathcal{N}(0,I)$ is a white noise
on $X$.
Note that $w:=\sqrt{\cC}\xi_0$ has covariance
operator $\cC$ and is hence a draw from $\mu_0$.
Recall that if $u$ is the current state
of the Markov chain, then $v$ is the proposal.
For simplicity we consider only Crank--Nicolson
proposals and not the MALA variants, so that
$\gamma=0$. However, the analysis generalises to the
Langevin proposals in a straightforward
fashion.

Rearranging (\ref{eqyarra}),
we see that the proposal $v$ satisfies
%
%e6.2 #&#
\begin{eqnarray}
\label{eqpropla} v&=& (I-\delta\theta\cK\cL )^{-1}\nonumber\\[-8pt]\\[-8pt]
&&{}\cdot \bigl( \bigl(I+
\delta(1-\theta) \cK\cL \bigr)u + \sqrt{2\delta\mathcal{K}} \xi_0
\bigr).\nonumber
\end{eqnarray}
If $\cK=I$, then the operator applied to $u$ is
bounded on $X$ for any $\theta\in(0,1]$.
If $\cK=\cC$, it is bounded for $\theta\in[0,1]$.
The white noise term is almost surely in
$X$ for
$\cK=I, \theta\in(0,1]$ and
$\cK=\cC, \theta\in[0,1]$.
The Crank--Nicolson proposal (\ref{eqone})
is found by letting $\mathcal{K} = I$ and
$\theta=\frac12$.
The preconditioned Crank--Nicolson proposal
(\ref{eqtwo}) is found by setting
$\mathcal{K} = \cC$ and $\theta=\frac12$.
The following theorem explains the choice $\theta=\frac12$.

%th6.2 #&#
\begin{theorem} \label{tac}
Let $\mu_0(X) = 1$, let $\Phi$ satisfy Assumption
\ref{ass1}(2) and assume that $\mu$ and $\mu_0$
are equivalent as measures with the Radon--Nikodym
derivative (\ref{eqratz2}).
Consider the proposal $v|u \sim q(u,\cdot)$ defined
by (\ref{eqpropla}) and the resulting
measure $\eta(du,dv)=q(u,dv)\mu(du)$ on $X \times X$.
For both $\cK=I$ and $\cK=\cC$
the measure $\eta^{\perp}=q(v,du)\mu(dv)$ is
equivalent to $\eta$ if and
only if $\theta=\frac12$. Furthermore, if $\theta=\frac12$, then
\[
\frac{d\eta^\perp}{d\eta}(u,v)= \exp \bigl(\Phi(u)-\Phi(v) \bigr).
\]
\end{theorem}

By use of the analysis of Metropolis--Hastings\break methods
on general state spaces\vadjust{\goodbreak} in \cite{Tie}, this theorem
shows that the Crank--Nicolson proposal (\ref{eqpropla})
leads to a well-defined MCMC algorithm in the function-space
setting, if and only if $\theta=\frac12$.
Note, relatedly, that the choice
$\theta= \frac{1}{2}$
has the\vspace*{1pt} desirable property that
$u \sim\cN(0,\cC)$ implies that $v \sim\cN(0,\cC)$:
thus, the prior measure is preserved under the proposal.
This mimics the behaviour of the SDE (\ref{eqspde})
for which the prior is an invariant measure. We have
thus justified the proposals (\ref{eqone})
and (\ref{eqtwo}) on function space. To
complete our analysis, it
remains to rule out the standard random walk
proposal (\ref{eqold}).

%th6.3 #&#
\begin{theorem} \label{lemsing}
Consider the proposal $v|u \sim\break q(u,\cdot)$ defined
by (\ref{eqold}) and the resulting
measure $\eta(du,dv)=q(u,dv)\mu(du)$ on $X \times X$.
For both $\cK=I$ and $\cK=\cC$
the measure $\eta^{\perp}=q(v,du)\mu(dv)$ is not
absolutely continuous with respect to $\eta$.
Thus, the MCMC method is not
defined on function space.
\end{theorem}

%s6.2 #&#
\subsection{The Acceptance Probability}
\label{ssecprob}

We now study the properties of the two Crank--Nicolson
methods with proposals (\ref{eqone}) and
(\ref{eqtwo}) in the limit $\delta\to0$,
showing that finite-dimensional intuition
carries over to this function space setting.
We define
\[
R(u,v)=\Phi(u)-\Phi(v)
\]
and note from (\ref{eqap}) that, for both
of the Crank--Nicolson proposals,
\[
a(u,v)=\min\bigl\{1,\exp \bigl(R(u,v) \bigr)\bigr\}.
\]
%
%The following theorem essentially shows that the function
%space random walk algorithms introduced in this paper
%behave similarly to their finite dimensional
%counterparts as the proposal variance $\delta$ approaches
%zero.

%th6.4 #&#
\begin{theorem} \label{thmthm}
Let $\mu_0$ be a Gaussian measure on a Hilbert space
$(X,\|\cdot\|)$ with $\mu_0(X)=1$
and let $\mu$ be an equivalent measure on $X$ given by
the Radon--Nikodym derivative (\ref{eqratz2}), satisfying
Assump-\break tions~\ref{ass1}(1) and~\ref{ass1}(2).
Then both the pCN and CN
algorithms with fixed $\delta$ are defined on $X$ and, furthermore,
the acceptance probability satisfies
\[
\lim_{\delta\to0} \bbE^{\eta} a(u,v)=1.
\]
\end{theorem}

%s6.3 #&#
\subsection{Scaling Limits and Spectral Gaps}
\label{ssecbrief}

There are two basic theories which have been
developed to explain the advantage of using
the algorithms introduced here which are
based on the function-space viewpoints.
The first is to prove scaling limits of the
algorithms, and the second is to establish spectral
gaps. The use of scaling limits was pioneered
for local-proposal Metropolis algorithms in the
papers \mbox{\cite{GGR97,RR98,RR01}}, and recently extended
to the hybrid Monte Carlo method \cite{besk10}.
All of this work concerned i.i.d. target distributions,
but recently it has been shown that the basic conclusions
of the theory, relating\vadjust{\goodbreak} to optimal scaling of proposal variance
with dimension, and optimal acceptance probability,
can be extended to the target measures of the form
(\ref{eqratz2}) which are central to this paper;
see \cite{MPS11,PST11a}. These results
show that the standard MCMC method must be scaled with proposal
variance (or time-step in the case of HMC) which
is inversely proportional to a power of $d_u$, the discretization
dimension, and that the number of steps required
grows under mesh refinement. The papers \cite{PST11,besk10b}
demonstrate that judicious modifications of these standard
algorithms, as described in this paper, lead to scaling limits
\emph{without} the need for scalings of proposal variance or
time-step which depend on dimension. These results indicate
that the number of steps required is stable under mesh refinement,
for these new methods, as demonstrated numerically in this paper.
The second approach, namely, the use of spectral gaps, offers
the opportunity to further substantiate these ideas: in
\cite{HSV12} it is shown that the pCN method has a dimension
independent spectral gap, whilst a standard random walk which
closely resembles it has spectral gap which shrinks with dimension.
This method of analysis, via spectral gaps, will be useful
for the analysis of many other MCMC algorithms arising in
high dimensions.

%s7 #&#
\section{Conclusions}
\label{secconc}

We have demonstrated the following points:

\begin{itemize}
\item A wide range of applications lead
naturally to problems defined via density with
respect to a Gaussian random field reference measure, or
variants on this structure.

\item Designing MCMC methods on function space,
and then discretizing the nonparametric problem, produces better insight
into algorithm design
than discretizing the nonparametric problem
and then applying standard MCMC methods.

\item The transferable idea underlying all the methods is that,
in the purely Gaussian case when only the reference measure
is sampled, the resulting MCMC method should accept with
probability one;
such methods may be identified by time-discretization
of certain stochastic dynamical systems which preserve the
Gaussian reference measure.

\item Using this methodology, we have highlighted new
random walk, Langevin and Hybrid Monte
Carlo Metropolis-type methods, appropriate
for problems where the posterior distribution
has density with respect to a Gaussian prior, all
of which can be implemented by means of small modifications
of existing codes.

\item We have applied these MCMC
methods to a range of problems,
demonstrating their efficacy in
comparison with standard methods,
and shown their flexibility with
respect to the incorporation of standard
ideas from MCMC technology such as Gibbs
sampling and estimation of noise precision
through conjugate Gamma priors.

\item We have pointed to the emerging body
of theoretical literature which substantiates the
desirable properties of the algorithms we have
highlighted here.
\end{itemize}

The ubiquity of Gaussian priors means that the
technology that is described in this article is
of immediate applicability to a wide range of
applications. The generality of the philosophy
that underlies our approach also suggests the possibility
of numerous further developments. In particular,
many existing algorithms can be modified to the
function space setting that is shown to be so desirable
here, when Gaussian priors underlie the desired target;
and many similar ideas can be expected to emerge
for the study of problems with non-Gaussian priors,
such as arise in wavelet based nonparametric estimation.

\section*{Acknowledgements}

S. L. Cotter is supported by EPSRC, ERC (FP7/\break2007-2013 and Grant
239870) and St. Cross College. G. O. Roberts is supported by EPSRC (especially
the CRiSM grant). A. M. Stuart is grateful to EPSRC, ERC and ONR for the
financial support of research which underpins this article. D. White is
supported by ERC.

%suskaldyti doi

% imsref loaded by lrinkeviciute, 2013-06-17 18:58:13
% imsref loaded by lrinkeviciute, 2013-06-17 19:20:14
% imsref loaded by lrinkeviciute, 2013-06-17 19:21:38


\begin{thebibliography}{67}
% BibTex style file: ims.bst, 2013-01-28
% Default style options (sort=0,type=number).
% Used options (sort=1,type=nameyear).

%b1 #&#
%  \bauthor{\bsnm{Sapatinas},~\bfnm{T.}\binits{T.}} \AND
%  \bauthor{\bsnm{Silverman},~\bfnm{B.~W.}\binits{B.~W.}}
%(\byear{1998}).

%b2 #&#
\bibitem{Mac10}
\begin{bmisc}[auto:STB|2013/06/05|13:45:01]
\bauthor{\bsnm{Adams},~\bfnm{R.~P.}\binits{R.~P.}},
  \bauthor{\bsnm{Murray},~\bfnm{I.}\binits{I.}} \AND
  \bauthor{\bsnm{Mackay},~\bfnm{D.~J.~C.}\binits{D.~J.~C.}}
(\byear{2009}).
\bhowpublished{The Gaussian process density sampler. In \textit{Advances in
  Neural Information Processing Systems 21}}.
\bptok{imsref}%
\end{bmisc}
\endbibitem

%b3 #&#
\bibitem{adler2}
\begin{bbook}[auto:STB|2013/06/05|13:45:01]
\bauthor{\bsnm{Adler},~\bfnm{R.~J.}\binits{R.~J.}}
(\byear{2010}).
\btitle{The Geometry of Random Fields}.
\bpublisher{SIAM}, \blocation{Philadeliphia, PA}.
\bptok{imsref}%
\end{bbook}
\endbibitem

%b4 #&#
\bibitem{ben02}
\begin{bbook}[mr]
\bauthor{\bsnm{Bennett},~\bfnm{Andrew~F.}\binits{A.~F.}}
(\byear{2002}).
\btitle{Inverse Modeling of the Ocean and Atmosphere}.
\bpublisher{Cambridge Univ. Press}, \blocation{Cambridge}.
\bid{doi={10.1017/CBO9780511535895}, mr={1920432}}
\bptok{imsref}%
\end{bbook}
\endbibitem

%b5 #&#
%  \bauthor{\bsnm{Roberts},~\bfnm{Gareth}\binits{G.}} \AND
%  \bauthor{\bsnm{Stuart},~\bfnm{Andrew}\binits{A.}}
%(\byear{2009}).
%  targets in high dimensions}.

%%b6 #&#
%  \bauthor{\bsnm{Stuart},~\bfnm{Andrew}\binits{A.}}
%(\byear{2009}).
%  dimensions}.
%In \bbooktitle{Monte {C}arlo and Quasi-{M}onte {C}arlo Methods 2008}
%(\beditor{\bfnm{P.}\binits{P.}~\bsnm{L'Ecuyer}} \AND
%  \beditor{\bfnm{Art~B.}\binits{A.~B.}~\bsnm{Owen}}, eds.)
%
%%b7 #&#
%  \bauthor{\bsnm{Stuart},~\bfnm{Andrew}\binits{A.}}
%(\byear{2009}).
%In \bbooktitle{I{CIAM} 07---6th {I}nternational {C}ongress on {I}ndustrial and
%  {A}pplied {M}athematics}

%b8 #&#
\bibitem{BRSV08}
\begin{barticle}[mr]
\bauthor{\bsnm{Beskos},~\bfnm{Alexandros}\binits{A.}},
  \bauthor{\bsnm{Roberts},~\bfnm{Gareth}\binits{G.}},
  \bauthor{\bsnm{Stuart},~\bfnm{Andrew}\binits{A.}} \AND
  \bauthor{\bsnm{Voss},~\bfnm{Jochen}\binits{J.}}
(\byear{2008}).
\btitle{M{CMC} methods for diffusion bridges}.
\bjournal{Stoch. Dyn.}
\bvolume{8}
\bpages{319--350}.
\bid{doi={10.1142/S0219493708002378}, issn={0219-4937}, mr={2444507}}
\bptok{imsref}%
\end{barticle}
\endbibitem

%b9 #&#
\bibitem{besk10b}
\begin{barticle}[mr]
\bauthor{\bsnm{Beskos},~\bfnm{A.}\binits{A.}},
  \bauthor{\bsnm{Pinski},~\bfnm{F.~J.}\binits{F.~J.}},
  \bauthor{\bsnm{Sanz-Serna},~\bfnm{J.~M.}\binits{J.~M.}} \AND
  \bauthor{\bsnm{Stuart},~\bfnm{A.~M.}\binits{A.~M.}}
(\byear{2011}).
\btitle{Hybrid {M}onte {C}arlo on {H}ilbert spaces}.
\bjournal{Stochastic Process. Appl.}
\bvolume{121}
\bpages{2201--2230}.
\bid{doi={10.1016/j.spa.2011.06.003}, issn={0304-4149}, mr={2822774}}
\bptok{imsref}%
\end{barticle}
\endbibitem

%b10 #&#
\bibitem{besk10}
\begin{bmisc}[auto:STB|2013/06/05|13:45:01]
\bauthor{\bsnm{Beskos},~\bfnm{A.}\binits{A.}},
  \bauthor{\bsnm{Pillai},~\bfnm{N.~S.}\binits{N.~S.}},
  \bauthor{\bsnm{Roberts},~\bfnm{G.~O.}\binits{G.~O.}},
  \bauthor{\bsnm{Sanz-Serna},~\bfnm{J.~M.}\binits{J.~M.}} \AND
  \bauthor{\bsnm{Stuart},~\bfnm{A.~M.}\binits{A.~M.}}
(\byear{2013}).
\bhowpublished{Optimal tuning of hybrid Monte Carlo. \textit{Bernoulli}. To
  appear. Available at \url{http://arxiv.org/abs/1001.4460}}.
\bptok{imsref}%
\end{bmisc}
\endbibitem

%b11 #&#
\bibitem{Co2008}
\begin{barticle}[mr]
\bauthor{\bsnm{Cotter},~\bfnm{C.~J.}\binits{C.~J.}}
(\byear{2008}).
\btitle{The variational particle-mesh method for matching curves}.
\bjournal{J. Phys. A}
\bvolume{41}
\bpages{344003, 18}.
\bid{doi={10.1088/1751-8113/41/34/344003}, issn={1751-8113}, mr={2456340}}
\bptok{imsref}%
\end{barticle}
\endbibitem

%b12 #&#
\bibitem{SC10}
\begin{bmisc}[auto:STB|2013/06/05|13:45:01]
\bauthor{\bsnm{Cotter},~\bfnm{S.~L.}\binits{S.~L.}}
(\byear{2010}).
\bhowpublished{Applications of MCMC methods on function spaces. Ph.D. thesis,
  Univ. Warwick}.
\bptok{imsref}%
\end{bmisc}
\endbibitem

%b13 #&#
\bibitem{CC10}
\begin{bmisc}[auto:STB|2013/06/05|13:45:01]
\bauthor{\bsnm{Cotter},~\bfnm{C.~J.}\binits{C.~J.}},
  \bauthor{\bsnm{Cotter},~\bfnm{S.~L.}\binits{S.~L.}} \AND
  \bauthor{\bsnm{Vialard},~\bfnm{F.~X.}\binits{F.~X.}}
(\byear{2013}).
\bhowpublished{Bayesian data assimilation in shape registration.
\textit{Inverse Problems}
\textbf{29} 045011}.
\bptok{imsref}%
\end{bmisc}
\endbibitem

%b14 #&#
\bibitem{cds11}
\begin{barticle}[mr]
\bauthor{\bsnm{Cotter},~\bfnm{S.~L.}\binits{S.~L.}},
  \bauthor{\bsnm{Dashti},~\bfnm{M.}\binits{M.}} \AND
  \bauthor{\bsnm{Stuart},~\bfnm{A.~M.}\binits{A.~M.}}
(\byear{2012}).
\btitle{Variational data assimilation using targetted random walks}.
\bjournal{Internat. J. Numer. Methods Fluids}
\bvolume{68}
\bpages{403--421}.
\bid{doi={10.1002/fld.2510}, issn={0271-2091}, mr={2880204}}
\bptok{imsref}%
\end{barticle}
\endbibitem

%b15 #&#
\bibitem{cdrs08}
\begin{barticle}[mr]
\bauthor{\bsnm{Cotter},~\bfnm{S.~L.}\binits{S.~L.}},
  \bauthor{\bsnm{Dashti},~\bfnm{M.}\binits{M.}},
  \bauthor{\bsnm{Robinson},~\bfnm{J.~C.}\binits{J.~C.}} \AND
  \bauthor{\bsnm{Stuart},~\bfnm{A.~M.}\binits{A.~M.}}
(\byear{2009}).
\btitle{Bayesian inverse problems for functions and applications to fluid
  mechanics}.
\bjournal{Inverse Problems}
\bvolume{25}
\bpages{115008, 43}.
\bid{doi={10.1088/0266-5611/25/11/115008}, issn={0266-5611}, mr={2558668}}
\bptok{imsref}%
\end{barticle}
\endbibitem

%b16 #&#
\bibitem{prato92}
\begin{bbook}[mr]
\bauthor{\bsnm{Da~Prato},~\bfnm{Giuseppe}\binits{G.}} \AND
  \bauthor{\bsnm{Zabczyk},~\bfnm{Jerzy}\binits{J.}}
(\byear{1992}).
\btitle{Stochastic Equations in Infinite Dimensions}.
\bseries{Encyclopedia of Mathematics and Its Applications}
\bvolume{44}.
\bpublisher{Cambridge Univ. Press}, \blocation{Cambridge}.
\bid{doi={10.1017/CBO9780511666223}, mr={1207136}}
\bptok{imsref}%
\end{bbook}
\endbibitem

%b17 #&#
\bibitem{ClDS10}
\begin{barticle}[mr]
\bauthor{\bsnm{Dashti},~\bfnm{Masoumeh}\binits{M.}},
  \bauthor{\bsnm{Harris},~\bfnm{Stephen}\binits{S.}} \AND
  \bauthor{\bsnm{Stuart},~\bfnm{Andrew}\binits{A.}}
(\byear{2012}).
\btitle{Besov priors for {B}ayesian inverse problems}.
\bjournal{Inverse Probl. Imaging}
\bvolume{6}
\bpages{183--200}.
\bid{doi={10.3934/ipi.2012.6.183}, issn={1930-8337}, mr={2942737}}
\bptnote{check year}%
\bptok{imsref}%
\end{barticle}
\endbibitem

%b18 #&#
%  \bauthor{\bsnm{Laird},~\bfnm{N.~M.}\binits{N.~M.}} \AND
%  \bauthor{\bsnm{Rubin},~\bfnm{D.~B.}\binits{D.~B.}}
%(\byear{1977}).

%b19 #&#
\bibitem{Dia88}
\begin{bincollection}[mr]
\bauthor{\bsnm{Diaconis},~\bfnm{Persi}\binits{P.}}
(\byear{1988}).
\btitle{Bayesian numerical analysis}.
In \bbooktitle{Statistical Decision Theory and Related Topics, {IV}, {V}ol.\ 1
  ({W}est {L}afayette, {I}nd., 1986)}
\bpages{163--175}.
\bpublisher{Springer}, \blocation{New York}.
\bid{mr={0927099}}
\bptok{imsref}%
\end{bincollection}
\endbibitem

%b20 #&#
\bibitem{duna87}
\begin{barticle}[auto:STB|2013/06/05|13:45:01]
\bauthor{\bsnm{Duane},~\bfnm{S.}\binits{S.}},
  \bauthor{\bsnm{Kennedy},~\bfnm{A.~D.}\binits{A.~D.}},
  \bauthor{\bsnm{Pendleton},~\bfnm{B.}\binits{B.}} \AND
  \bauthor{\bsnm{Roweth},~\bfnm{D.}\binits{D.}}
(\byear{1987}).
\btitle{Hybrid Monte Carlo}.
\bjournal{Phys. Lett. B}
\bvolume{195}
\bpages{216--222}.
\bptok{imsref}%
\end{barticle}
\endbibitem

%b21 #&#
\bibitem{giro11}
\begin{barticle}[mr]
\bauthor{\bsnm{Girolami},~\bfnm{Mark}\binits{M.}} \AND
  \bauthor{\bsnm{Calderhead},~\bfnm{Ben}\binits{B.}}
(\byear{2011}).
\btitle{Riemann manifold {L}angevin and {H}amiltonian {M}onte {C}arlo methods
  (with discussion)}.
\bjournal{J. R. Stat. Soc. Ser. B Stat. Methodol.}
\bvolume{73}
\bpages{123--214}.
\bid{doi={10.1111/j.1467-9868.2010.00765.x}, issn={1369-7412}, mr={2814492}}
\bptok{imsref}%
\end{barticle}
\endbibitem

%b22 #&#
\bibitem{GlTrYo04}
\begin{bincollection}[auto:STB|2013/06/05|13:45:01]
\bauthor{\bsnm{Glaunes},~\bfnm{J.}\binits{J.}},
  \bauthor{\bsnm{Trouv{\'e}},~\bfnm{A.}\binits{A.}} \AND
  \bauthor{\bsnm{Younes},~\bfnm{L.}\binits{L.}}
(\byear{2004}).
\btitle{Diffeomorphic matching of distributions: A new approach for unlabelled
  point-sets and sub-manifolds matching}.
In \bbooktitle{Computer Vision and Pattern Recognition, 2004. CVPR 2004. Proceedings of the 2004 IEEE
Computer Society Conference on 2}
\bpages{712--718}.
\bpublisher{IEEE}.
\bptok{imsref}%
\end{bincollection}
\endbibitem

%b23 #&#
\bibitem{HSV05}
\begin{barticle}[mr]
\bauthor{\bsnm{Hairer},~\bfnm{M.}\binits{M.}},
  \bauthor{\bsnm{Stuart},~\bfnm{A.~M.}\binits{A.~M.}} \AND
  \bauthor{\bsnm{Voss},~\bfnm{J.}\binits{J.}}
(\byear{2007}).
\btitle{Analysis of {SPDE}s arising in path sampling. {II}. {T}he nonlinear
  case}.
\bjournal{Ann. Appl. Probab.}
\bvolume{17}
\bpages{1657--1706}.
\bid{doi={10.1214/07-AAP441}, issn={1050-5164}, mr={2358638}}
\bptok{imsref}%
\end{barticle}
\endbibitem

%b24 #&#
\bibitem{HSV09}
\begin{bincollection}[mr]
\bauthor{\bsnm{Hairer},~\bfnm{Martin}\binits{M.}},
  \bauthor{\bsnm{Stuart},~\bfnm{Andrew}\binits{A.}} \AND
  \bauthor{\bsnm{Vo{\ss}},~\bfnm{Jochen}\binits{J.}}
(\byear{2009}).
\btitle{Sampling conditioned diffusions}.
In \bbooktitle{Trends in Stochastic Analysis}.
\bseries{London Mathematical Society Lecture Note Series}
\bvolume{353}
\bpages{159--185}.
\bpublisher{Cambridge Univ. Press}, \blocation{Cambridge}.
\bid{mr={2562154}}
\bptok{imsref}%
\end{bincollection}
\endbibitem

%b25 #&#
\bibitem{HSV10}
\begin{bincollection}[mr]
\bauthor{\bsnm{Hairer},~\bfnm{M.}\binits{M.}},
  \bauthor{\bsnm{Stuart},~\bfnm{A.}\binits{A.}} \AND
  \bauthor{\bsnm{Voss},~\bfnm{J.}\binits{J.}}
(\byear{2011}).
\btitle{Signal processing problems on function space: {B}ayesian formulation,
  stochastic {PDE}s and effective {MCMC} methods}.
In \bbooktitle{The {O}xford Handbook of Nonlinear Filtering}
(\beditor{D. Crisan} and \beditor{B. Rozovsky}, eds.)
\bpages{833--873}.
\bpublisher{Oxford Univ. Press}, \blocation{Oxford}.
\bid{mr={2884617}}
\bptnote{check year}%
\bptok{imsref}%
\end{bincollection}
\endbibitem

%b26 #&#
%  \bauthor{\bsnm{Stuart},~\bfnm{A.}\binits{A.}} \AND
%  \bauthor{\bsnm{Voss},~\bfnm{J.}\binits{J.}}
%(\byear{2011}).
%  stochastic {PDE}s and effective {MCMC} methods}.
%In \bbooktitle{The {O}xford Handbook of Nonlinear Filtering}

%b27 #&#
\bibitem{HSV12}
\begin{bmisc}[auto:STB|2013/06/05|13:45:01]
\bauthor{\bsnm{Hairer},~\bfnm{M.}\binits{M.}},
  \bauthor{\bsnm{Stuart},~\bfnm{A.~M.}\binits{A.~M.}} \AND
  \bauthor{\bsnm{Vollmer},~\bfnm{S.}\binits{S.}}
(\byear{2013}).
\bhowpublished{Spectral gaps for a Metropolis--{H}astings algorithm in infinite
  dimensions. Available at
  \texttt{\href{http://arxiv.org/abs/1112.1392}{http://}
  \href{http://arxiv.org/abs/1112.1392}{arxiv.org/abs/1112.1392}}}.
\bptok{imsref}%
\end{bmisc}
\endbibitem

%b28 #&#
\bibitem{HSWV05}
\begin{barticle}[mr]
\bauthor{\bsnm{Hairer},~\bfnm{M.}\binits{M.}},
  \bauthor{\bsnm{Stuart},~\bfnm{A.~M.}\binits{A.~M.}},
  \bauthor{\bsnm{Voss},~\bfnm{J.}\binits{J.}} \AND
  \bauthor{\bsnm{Wiberg},~\bfnm{P.}\binits{P.}}
(\byear{2005}).
\btitle{Analysis of {SPDE}s arising in path sampling. {I}. {T}he {G}aussian
  case}.
\bjournal{Commun. Math. Sci.}
\bvolume{3}
\bpages{587--603}.
\bid{issn={1539-6746}, mr={2188686}}
\bptok{imsref}%
\end{barticle}
\endbibitem

%b29 #&#
%(\byear{1970}).
%  applications}.

%b30 #&#
\bibitem{hillssmith}
\begin{bincollection}[mr]
\bauthor{\bsnm{Hills},~\bfnm{Susan~E.}\binits{S.~E.}} \AND
  \bauthor{\bsnm{Smith},~\bfnm{Adrian F.~M.}\binits{A.~F.~M.}}
(\byear{1992}).
\btitle{Parameterization issues in {B}ayesian inference}.
In \bbooktitle{Bayesian Statistics, 4 ({P}e\~n\'\i Scola, 1991)}
\bpages{227--246}.
\bpublisher{Oxford Univ. Press}, \blocation{New York}.
\bid{mr={1380279}}
\bptok{imsref}%
\end{bincollection}
\endbibitem

%b31 #&#
\bibitem{hj10}
\begin{bbook}[mr]
\beditor{\bsnm{Hjort},~\bfnm{Nils~Lid}\binits{N.~L.}},
  \beditor{\bsnm{Holmes},~\bfnm{Chris}\binits{C.}},
  \beditor{\bsnm{M{\"u}ller},~\bfnm{Peter}\binits{P.}} \AND
  \beditor{\bsnm{Walker},~\bfnm{Stephen~G.}\binits{S.~G.}}, eds.
(\byear{2010}).
\btitle{Bayesian Nonparametrics}.
\bseries{Cambridge Series in Statistical and Probabilistic Mathematics}
\bvolume{28}.
\bpublisher{Cambridge Univ. Press}, \blocation{Cambridge}.
\bid{doi={10.1017/CBO9780511802478}, mr={2722987}}
\bptok{imsref}%
\end{bbook}
\endbibitem

%b32 #&#
\bibitem{iserles04}
\begin{bbook}[auto:STB|2013/06/05|13:45:01]
\bauthor{\bsnm{Iserles},~\bfnm{A.}\binits{A.}}
(\byear{2004}).
\btitle{A First Course in the Numerical Analysis of Differential Equations}.
\bpublisher{Cambridge Univ. Press}, \blocation{Cambridge}.
\bptok{imsref}%
\end{bbook}
\endbibitem

%b33 #&#
\bibitem{kal03}
\begin{bbook}[auto:STB|2013/06/05|13:45:01]
\bauthor{\bsnm{Kalnay},~\bfnm{E.}\binits{E.}}
(\byear{2003}).
\btitle{Atmospheric Modeling, Data Assimilation and Predictability}.
\bpublisher{Cambridge Univ. Press}, \blocation{Cambridge}.
\bptok{imsref}%
\end{bbook}
\endbibitem

%b34 #&#
%  \bauthor{\bsnm{Simard},~\bfnm{Richard}\binits{R.}}
%(\byear{2007}).
%  generators}.

%b35 #&#
\bibitem{lemm03}
\begin{bbook}[mr]
\bauthor{\bsnm{Lemm},~\bfnm{J{\"o}rg~C.}\binits{J.~C.}}
(\byear{2003}).
\btitle{Bayesian Field Theory}.
\bpublisher{Johns Hopkins Univ. Press}, \blocation{Baltimore, MD}.
\bid{mr={1987925}}
\bptok{imsref}%
\end{bbook}
\endbibitem

%b36 #&#
\bibitem{Liu}
\begin{bbook}[mr]
\bauthor{\bsnm{Liu},~\bfnm{Jun~S.}\binits{J.~S.}}
(\byear{2001}).
\btitle{Monte {C}arlo Strategies in Scientific Computing}.
\bpublisher{Springer}, \blocation{New York}.
\bid{mr={1842342}}
\bptok{imsref}%
\end{bbook}
\endbibitem

%b37 #&#
\bibitem{MPS11}
\begin{barticle}[mr]
\bauthor{\bsnm{Mattingly},~\bfnm{Jonathan~C.}\binits{J.~C.}},
  \bauthor{\bsnm{Pillai},~\bfnm{Natesh~S.}\binits{N.~S.}} \AND
  \bauthor{\bsnm{Stuart},~\bfnm{Andrew~M.}\binits{A.~M.}}
(\byear{2012}).
\btitle{Diffusion limits of the random walk {M}etropolis algorithm in high
  dimensions}.
\bjournal{Ann. Appl. Probab.}
\bvolume{22}
\bpages{881--930}.
\bid{doi={10.1214/10-AAP754}, issn={1050-5164}, mr={2977981}}
\bptok{imsref}%
\end{barticle}
\endbibitem

%b38 #&#
\bibitem{mct}
\begin{barticle}[auto:STB|2013/06/05|13:45:01]
\bauthor{\bsnm{McLaughlin},~\bfnm{D.}\binits{D.}} \AND
  \bauthor{\bsnm{Townley},~\bfnm{L.~R.}\binits{L.~R.}}
(\byear{1996}).
\btitle{A reassessment of the groundwater inverse problem}.
\bjournal{Water Res. Res.}
\bvolume{32}
\bpages{1131--1161}.
\bptok{imsref}%
\end{barticle}
\endbibitem

%%b39 #&#
%  \bauthor{\bparticle{van} \bsnm{Dyk},~\bfnm{David}\binits{D.}}
%(\byear{1997}).
%
%%b40 #&#
%  \bauthor{\bsnm{Rosenbluth},~\bfnm{R.~W.}\binits{R.~W.}},
%  \bauthor{\bsnm{Teller},~\bfnm{M.~N.}\binits{M.~N.}} \AND
%  \bauthor{\bsnm{Teller},~\bfnm{E.}\binits{E.}}
%(\byear{1953}).

%b41 #&#
\bibitem{MiYo2001}
\begin{barticle}[auto:STB|2013/06/05|13:45:01]
\bauthor{\bsnm{Miller},~\bfnm{M.~T.}\binits{M.~T.}} \AND
  \bauthor{\bsnm{Younes},~\bfnm{L.}\binits{L.}}
(\byear{2001}).
\btitle{Group actions, homeomorphisms, and matching: A general framework}.
\bjournal{Int. J. Comput. Vis.}
\bvolume{41}
\bpages{61--84}.
\bptok{imsref}%
\end{barticle}
\endbibitem

%b42 #&#
\bibitem{neal96}
\begin{bbook}[auto:STB|2013/06/05|13:45:01]
\bauthor{\bsnm{Neal},~\bfnm{R.~M.}\binits{R.~M.}}
(\byear{1996}).
\btitle{Bayesian Learning for Neural Networks}.
\bpublisher{Springer}, \blocation{New York}.
\bptok{imsref}%
\end{bbook}
\endbibitem

%b43 #&#
\bibitem{Neal98}
\begin{bmisc}[mr]
\bauthor{\bsnm{Neal},~\bfnm{Radford~M.}\binits{R.~M.}}
(\byear{1998}).
\bhowpublished{Regression and classification using {G}aussian process priors.
  Available at \texttt{\href{http://www.cs.toronto.edu/\textasciitilde radford/valencia.abstract.html}{http://www.cs.toronto.edu/\textasciitilde radford/valencia.}
  \href{http://www.cs.toronto.edu/\textasciitilde radford/valencia.abstract.html}{abstract.html}}}.
\bptok{imsref}%
\end{bmisc}
\endbibitem

%b44 #&#
\bibitem{neal2010}
\begin{bincollection}[mr]
\bauthor{\bsnm{Neal},~\bfnm{Radford~M.}\binits{R.~M.}}
(\byear{2011}).
\btitle{M{CMC} using {H}amiltonian dynamics}.
In \bbooktitle{Handbook of {M}arkov Chain {M}onte {C}arlo}
\bpages{113--162}.
\bpublisher{CRC Press}, \blocation{Boca Raton, FL}.
\bid{mr={2858447}}
\bptnote{check year}%
\bptok{imsref}%
\end{bincollection}
\endbibitem

%b45 #&#
\bibitem{hag99}
\begin{bincollection}[mr]
\bauthor{\bsnm{O'Hagan},~\bfnm{Anthony}\binits{A.}},
  \bauthor{\bsnm{Kennedy},~\bfnm{Marc~C.}\binits{M.~C.}} \AND
  \bauthor{\bsnm{Oakley},~\bfnm{Jeremy~E.}\binits{J.~E.}}
(\byear{1999}).
\btitle{Uncertainty analysis and other inference tools for complex computer
  codes}.
In \bbooktitle{Bayesian Statistics, 6 ({A}lcoceber, 1998)}
\bpages{503--524}.
\bpublisher{Oxford Univ. Press}, \blocation{New York}.
\bid{mr={1724872}}
\bptnote{check related}%
\bptok{imsref}%
\end{bincollection}
\endbibitem

%b46 #&#
\bibitem{PST11a}
\begin{barticle}[mr]
\bauthor{\bsnm{Pillai},~\bfnm{Natesh~S.}\binits{N.~S.}},
  \bauthor{\bsnm{Stuart},~\bfnm{Andrew~M.}\binits{A.~M.}} \AND
  \bauthor{\bsnm{Thi{\'e}ry},~\bfnm{Alexandre~H.}\binits{A.~H.}}
(\byear{2012}).
\btitle{Optimal scaling and diffusion limits for the {L}angevin algorithm in
  high dimensions}.
\bjournal{Ann. Appl. Probab.}
\bvolume{22}
\bpages{2320--2356}.
\bid{issn={1050-5164}, mr={3024970}}
\bptok{imsref}%
\end{barticle}
\endbibitem

%b47 #&#
\bibitem{PST11}
\begin{bmisc}[auto:STB|2013/06/05|13:45:01]
\bauthor{\bsnm{Pillai},~\bfnm{N.~S.}\binits{N.~S.}},
  \bauthor{\bsnm{Stuart},~\bfnm{A.~M.}\binits{A.~M.}} \AND
  \bauthor{\bsnm{Thiery},~\bfnm{A.~H.}\binits{A.~H.}}
(\byear{2012}).
\bhowpublished{On the random walk Metropolis algorithm for Gaussian random
  field priors and gradient flow. Available at
  \url{http://arxiv.org/abs/1108.1494}}.
\bptok{imsref}%
\end{bmisc}
\endbibitem

%%b48 #&#
%  \bauthor{\bsnm{Teukolsky},~\bfnm{Saul~A.}\binits{S.~A.}},
%  \bauthor{\bsnm{Vetterling},~\bfnm{William~T.}\binits{W.~T.}} \AND
%  \bauthor{\bsnm{Flannery},~\bfnm{Brian~P.}\binits{B.~P.}}
%(\byear{2002}).
%The Art of Scientific Computing},
%
%%b49 #&#
%  \bauthor{\bsnm{Silverman},~\bfnm{B.~W.}\binits{B.~W.}}
%(\byear{2005}).

%b50 #&#
\bibitem{RM67}
\begin{bbook}[auto:STB|2013/06/05|13:45:01]
\bauthor{\bsnm{Richtmyer},~\bfnm{D.}\binits{D.}} \AND
  \bauthor{\bsnm{Morton},~\bfnm{K.~W.}\binits{K.~W.}}
(\byear{1967}).
\btitle{Difference Methods for Initial Value Problems}.
\bpublisher{Wiley}, \blocation{New York}.
\bptok{imsref}%
\end{bbook}
\endbibitem

%b51 #&#
\bibitem{RC99}
\begin{bbook}[mr]
\bauthor{\bsnm{Robert},~\bfnm{Christian~P.}\binits{C.~P.}} \AND
  \bauthor{\bsnm{Casella},~\bfnm{George}\binits{G.}}
(\byear{1999}).
\btitle{Monte {C}arlo Statistical Methods}.
\bpublisher{Springer}, \blocation{New York}.
\bid{mr={1707311}}
\bptok{imsref}%
\end{bbook}
\endbibitem

%b52 #&#
\bibitem{GGR97}
\begin{barticle}[mr]
\bauthor{\bsnm{Roberts},~\bfnm{G.~O.}\binits{G.~O.}},
  \bauthor{\bsnm{Gelman},~\bfnm{A.}\binits{A.}} \AND
  \bauthor{\bsnm{Gilks},~\bfnm{W.~R.}\binits{W.~R.}}
(\byear{1997}).
\btitle{Weak convergence and optimal scaling of random walk {M}etropolis
  algorithms}.
\bjournal{Ann. Appl. Probab.}
\bvolume{7}
\bpages{110--120}.
\bid{doi={10.1214/aoap/1034625254}, issn={1050-5164}, mr={1428751}}
\bptok{imsref}%
\end{barticle}
\endbibitem

%b53 #&#
\bibitem{RR98}
\begin{barticle}[mr]
\bauthor{\bsnm{Roberts},~\bfnm{Gareth~O.}\binits{G.~O.}} \AND
  \bauthor{\bsnm{Rosenthal},~\bfnm{Jeffrey~S.}\binits{J.~S.}}
(\byear{1998}).
\btitle{Optimal scaling of discrete approximations to {L}angevin diffusions}.
\bjournal{J.~R. Stat. Soc. Ser. B Stat. Methodol.}
\bvolume{60}
\bpages{255--268}.
\bid{doi={10.1111/1467-9868.00123}, issn={1369-7412}, mr={1625691}}
\bptok{imsref}%
\end{barticle}
\endbibitem

%b54 #&#
\bibitem{RR01}
\begin{barticle}[mr]
\bauthor{\bsnm{Roberts},~\bfnm{Gareth~O.}\binits{G.~O.}} \AND
  \bauthor{\bsnm{Rosenthal},~\bfnm{Jeffrey~S.}\binits{J.~S.}}
(\byear{2001}).
\btitle{Optimal scaling for various {M}etropolis--{H}astings algorithms}.
\bjournal{Statist. Sci.}
\bvolume{16}
\bpages{351--367}.
\bid{doi={10.1214/ss/1015346320}, issn={0883-4237}, mr={1888450}}
\bptok{imsref}%
\end{barticle}
\endbibitem

%b55 #&#
\bibitem{RS01}
\begin{barticle}[mr]
\bauthor{\bsnm{Roberts},~\bfnm{G.~O.}\binits{G.~O.}} \AND
  \bauthor{\bsnm{Stramer},~\bfnm{O.}\binits{O.}}
(\byear{2001}).
\btitle{On inference for partially observed nonlinear diffusion models using
  the {M}etropolis--{H}astings algorithm}.
\bjournal{Biometrika}
\bvolume{88}
\bpages{603--621}.
\bid{doi={10.1093/biomet/88.3.603}, issn={0006-3444}, mr={1859397}}
\bptok{imsref}%
\end{barticle}
\endbibitem

%b56 #&#
\bibitem{robtwe96}
\begin{barticle}[mr]
\bauthor{\bsnm{Roberts},~\bfnm{Gareth~O.}\binits{G.~O.}} \AND
  \bauthor{\bsnm{Tweedie},~\bfnm{Richard~L.}\binits{R.~L.}}
(\byear{1996}).
\btitle{Exponential convergence of {L}angevin distributions and their discrete
  approximations}.
\bjournal{Bernoulli}
\bvolume{2}
\bpages{341--363}.
\bid{doi={10.2307/3318418}, issn={1350-7265}, mr={1440273}}
\bptok{imsref}%
\end{barticle}
\endbibitem

%b57 #&#
\bibitem{RH05}
\begin{bbook}[mr]
\bauthor{\bsnm{Rue},~\bfnm{H{\aa}vard}\binits{H.}} \AND
  \bauthor{\bsnm{Held},~\bfnm{Leonhard}\binits{L.}}
(\byear{2005}).
\btitle{Gaussian {M}arkov Random Fields:
Theory and Applications}.
\bseries{Monographs on Statistics and Applied Probability}
\bvolume{104}.
\bpublisher{Chapman \& Hall/CRC}, \blocation{Boca Raton, FL}.
\bid{doi={10.1201/9780203492024}, mr={2130347}}
\bptok{imsref}%
\end{bbook}
\endbibitem

%b58 #&#
\bibitem{SmRo}
\begin{barticle}[mr]
\bauthor{\bsnm{Smith},~\bfnm{A.~F.~M.}\binits{A.~F.~M.}} \AND
  \bauthor{\bsnm{Roberts},~\bfnm{G.~O.}\binits{G.~O.}}
(\byear{1993}).
\btitle{Bayesian computation via the {G}ibbs sampler and related {M}arkov chain
  {M}onte {C}arlo methods}.
\bjournal{J. R. Stat. Soc. Ser. B Stat. Methodol.}
\bvolume{55}
\bpages{3--23}.
\bid{issn={0035-9246}, mr={1210421}}
\bptok{imsref}%
\end{barticle}
\endbibitem

%b59 #&#
\bibitem{sokal}
\begin{bmisc}[auto:STB|2013/06/05|13:45:01]
\bauthor{\bsnm{Sokal},~\bfnm{A.~D.}\binits{A.~D.}}
(\byear{1989}).
\bhowpublished{Monte Carlo methods in statistical mechanics: Foundations and
  new algorithms, Univ. Lausanne, B\^atiment des sciences de physique
  Troisi\`eme Cycle de la physique en Suisse romande}.
\bptok{imsref}%
\end{bmisc}
\endbibitem

%b60 #&#
\bibitem{st99}
\begin{bbook}[mr]
\bauthor{\bsnm{Stein},~\bfnm{Michael~L.}\binits{M.~L.}}
(\byear{1999}).
\btitle{Interpolation of Spatial Data:
Some Theory for Kriging}.
\bpublisher{Springer}, \blocation{New York}.
\bid{doi={10.1007/978-1-4612-1494-6}, mr={1697409}}
\bptok{imsref}%
\end{bbook}
\endbibitem

%b61 #&#
\bibitem{Stuart10}
\begin{barticle}[mr]
\bauthor{\bsnm{Stuart},~\bfnm{A.~M.}\binits{A.~M.}}
(\byear{2010}).
\btitle{Inverse problems: A {B}ayesian perspective}.
\bjournal{Acta Numer.}
\bvolume{19}
\bpages{451--559}.
\bid{doi={10.1017/S0962492910000061}, issn={0962-4929}, mr={2652785}}
\bptok{imsref}%
\end{barticle}
\endbibitem

%b62 #&#
\bibitem{SVW04}
\begin{barticle}[mr]
\bauthor{\bsnm{Stuart},~\bfnm{Andrew~M.}\binits{A.~M.}},
  \bauthor{\bsnm{Voss},~\bfnm{Jochen}\binits{J.}} \AND
  \bauthor{\bsnm{Wiberg},~\bfnm{Petter}\binits{P.}}
(\byear{2004}).
\btitle{Fast communication conditional path sampling of {SDE}s and the
  {L}angevin {MCMC} method}.
\bjournal{Commun. Math. Sci.}
\bvolume{2}
\bpages{685--697}.
\bid{issn={1539-6746}, mr={2119934}}
\bptok{imsref}%
\end{barticle}
\endbibitem

%%b63 #&#
%  \bauthor{\bsnm{Wong},~\bfnm{Wing~Hung}\binits{W.~H.}}
%(\byear{1987}).

%b64 #&#
\bibitem{Tie}
\begin{barticle}[mr]
\bauthor{\bsnm{Tierney},~\bfnm{Luke}\binits{L.}}
(\byear{1998}).
\btitle{A note on {M}etropolis--{H}astings kernels for general state spaces}.
\bjournal{Ann. Appl. Probab.}
\bvolume{8}
\bpages{1--9}.
\bid{doi={10.1214/aoap/1027961031}, issn={1050-5164}, mr={1620401}}
\bptok{imsref}%
\end{barticle}
\endbibitem

%b65 #&#
\bibitem{VaGl2005}
\begin{bmisc}[auto:STB|2013/06/05|13:45:01]
\bauthor{\bsnm{Vaillant},~\bfnm{M.}\binits{M.}} \AND
  \bauthor{\bsnm{Glaunes},~\bfnm{J.}\binits{J.}}
(\byear{2005}).
\bhowpublished{Surface matching via currents.
In \textit{Information Processing in Medical Imaging}
381--392. Springer, Berlin\phantom{}}.
\bptok{imsref}%
\end{bmisc}
\endbibitem

\bibitem{van2013reversible}
\begin{bmisc}[mr]
\bauthor{\bsnm{van der Meulen},~\bfnm{Frank}\binits{F.}},
\bauthor{\bsnm{Schauer},~\bfnm{Moritz}\binits{M.}} \AND
\bauthor{\bsnm{van Zanten},~\bfnm{Harry}\binits{H.}}
(\byear{2013}).
\bhowpublished{Reversible jump MCMC for nonparametric drift estimation for diffusion
processes. \textit{Comput. Statist. Data Anal.} To appear}.
\bptok{imsref}%
\end{bmisc}
\endbibitem

%b66 #&#
\bibitem{zhao2000bayesian}
\begin{barticle}[mr]
\bauthor{\bsnm{Zhao},~\bfnm{Linda~H.}\binits{L.~H.}}
(\byear{2000}).
\btitle{Bayesian aspects of some nonparametric problems}.
\bjournal{Ann. Statist.}
\bvolume{28}
\bpages{532--552}.
\bid{doi={10.1214/aos/1016218229}, issn={0090-5364}, mr={1790008}}
\bptok{imsref}%
\end{barticle}
\endbibitem

\end{thebibliography}
\end{document}